\documentclass[envcountsame]{llncs}

\usepackage[english]{babel}
\usepackage[T1]{fontenc}
\usepackage[utf8]{inputenc}
\usepackage{xspace}
\usepackage{amsmath,amsfonts,amssymb}

\usepackage{graphicx}

\usepackage[amsmath]{ntheorem}
\usepackage{cleveref}

\usepackage{algorithm}
\usepackage{algpseudocode}
\algnotext{EndFor}
\algnotext{EndIf}
\algnotext{EndProcedure}
\algnotext{EndFunction}
\algnotext{EndWhile}

\usepackage{cite}

\usepackage{algorithm}
\usepackage{algpseudocode}

\usepackage{placeins}

\usepackage{subfig}

\newcommand{\cp}{\mathbf{P}\xspace}
\newcommand{\cnp}{\mathbf{NP}\xspace}

\newcommand{\QQ}{\mathbb{Q}}
\newcommand{\NN}{\mathbb{N}}
\newcommand{\nil}{\mathbf{nil}}

\newcommand{\vv}{\mathbf{v}}

\begin{document}

\pagestyle{headings}

\mainmatter

\title{Tree-Deletion Pruning in Label-Correcting Algorithms for the Multiobjective Shortest Path Problem}

\author{Fritz B{\"o}kler \and Petra Mutzel}

\institute{Department of Computer Science, TU Dortmund, Germany\\
\email{\{fritz.boekler, petra.mutzel\}@tu-dortmund.de}}

\maketitle

\begin{abstract}
In this paper, we re-evaluate the basic strategies for label correcting algorithms for the multiobjective shortest path (MOSP) problem, i.e., node and label selection.
In contrast to common believe, we show that---when carefully implemented---the node-selection strategy usually beats the label-selection strategy.
Moreover, we present a new pruning method which is easy to implement and performs very well on real-world road networks.
In this study, we test our hypotheses on artificial MOSP instances from the literature with up to 15 objectives and real-world road networks with up to almost 160,000 nodes.
\end{abstract}

\section{Introduction}

In this paper we are concerned with one of the most famous problems from multiobjective optimization, the multiobjective shortest path (MOSP) problem.
We are given a directed graph $G$, consisting of a finite set of nodes $V$ and a set of directed arcs $A \subseteq V \times V$.
We are interested in paths between a given \emph{source node} $s$ and a given \emph{target node} $t$.
Instead of a single-objective cost function, we are given an objective function $c$ which maps each arc to a vector, i.e., $c: A \rightarrow \QQ^d$ for $d \in \NN$.
The set of all directed paths from $s$ to $t$ in a given graph is called $\mathcal{P}_{s,t}$ and we assume that the objective function $c$ is extended on these paths in the canonical way.

In contrast to the single-objective case, where there exists only one unique optimal value, in the multiobjective case there usually does not exist a path minimizing all objectives at once.
Thus, we are concerned with finding the \emph{Pareto-front} of all $s$-$t$-paths, i.e., the minimal vectors of the set $c(\mathcal{P}_{s,t})$ with respect to the canonical componentwise partial order on vectors $\leq$.
Moreover, we also want to find for each point $y$ of the Pareto-front one representative path $p\in \mathcal{P}_{s,t}$, such that $c(p) = y$.
Each such path is called \emph{Pareto-optimal}.
For more information on multiobjective path and tree problems we refer the reader to the latest survey \cite{CP12}.

It is long known that the Pareto-front of a MOSP instance can be of exponential size in the input \cite{H79}.
Moreover, it has been recently shown in \cite{BEMM16}, that there does not exist an output-sensitive algorithm for this problem even in the case of $d=2$ unless $\cp = \cnp$.

\subsection{Previous Work}
The techniques used for solving the MOSP problem are based on labeling algorithms.
The majority of the literature is concerned with the biobjective case.
The latest computational study for more than 2 objectives is from 2009 \cite{PS09} and compares 27 variants of labeling algorithms on 9,050 artificial instances.
These are the instances we also use for our study.
In summary, a label correcting version with a label-selection strategy in a FIFO manner is concluded to be the fastest strategy on the instance classes provided.
The authors do not investigate a node-selection strategy with the argument that it is harder to implement and is less efficient (cf. also \cite{PS07}).

In an older study from 2001 \cite{GM01}, also label-selection and node-selection strategies are compared.
The authors conclude that, in general, label-selection methods are faster than node-selection methods.
However, the test set is rather small, consisting of only 8 artificial grid-graph instances ranging from 100 to 500 nodes and 2 to 4 objectives and 18 artificial random-graph instances ranging from 500 to 40,000 nodes and densities of 1.5 to 30 with 2 to 4 objectives.

In the work by Delling and Wagner \cite{DW09}, the authors solve a variant of the multiobjective shortest path problem where a preprocessing is allowed and we want to query the Pareto-front of paths between a pair of nodes as fast as possible.
The authors use a variant of SHARC to solve this problem.
Though being a different problem, this study is the first computational study where an implementation is tested on real-world road networks instead of artificial instances.
The instances have sizes of 30,661, 71,619 and 892,392 nodes and 2 to 4 objectives.
Though, the largest instance could only be solved using highly correlated objective functions resulting in Pareto-front sizes of only at most 2.5 points on average.

\subsection{Our Contribution}

In this paper, we investigate the efficiency of label correcting methods for the multiobjective shortest path problem.
We focus on label correcting methods, because the literature (cf. \cite{PS09, RE09}) and our experience shows that label setting algorithms do not perform well on instances with more than two objectives.
We investigate the question if label-selection or node-selection methods are more promising and test codes based on the recent literature.

We also perform the first computational study of these algorithms not only on artificial instances but also on real-world road networks based on the road network of Western Europe provided by the PTV AG for scientific use.
The road network sizes vary from 23,094 to 159,945 nodes and include three objective functions.
The artificial instances are taken from the latest study on the MOSP problem.

Moreover, we propose a new pruning technique which performs very well on the road networks, achieving large speed-ups.
We also show the limits of this technique and reason under which circumstances it works well.

\subsection{Organization}

In Sec. \ref{sec:labeling}, we describe the basic techniques of labeling algorithms in multiobjective optimization.
The tree-deletion pruning is the concern of Sec. \ref{sec:td}.
Because the implementation of the algorithms is crucial for this computational study, we give details in Sec. \ref{sec:impl}.
The computational study and results then follow in Sec. \ref{sec:study}.

\section{Multiobjective Labeling Algorithms}\label{sec:labeling}

In general, a labeling algorithm for enumerating the Pareto-front of paths in a graph maintains a set of labels $L_u$ at each node $u\in V$.
A label is a tuple which holds its cost vector, the associated node and, for retrieving the actual path, a reference to the predecessor label.
The algorithm is initialized by setting each label-set to $\emptyset$ and adding the label $(\mathbf{0}, s, \nil)$ to $L_s$.

These algorithms can be divided into two groups, depending if they select either a label or a node in each iteration.
These strategies are called \emph{label} or \emph{node selection strategies}, respectively.
When we select a label $\ell = (\mathbf{v}, u, \hat{\ell})$ at a node $u$, this label is \emph{pushed} along all out-arcs $a = (u,w)$ of $u$, meaning that a new label at the head of the arc is created with cost $\mathbf{v} + c(a)$, predecessor label $\ell$ and associated node $w$.
This strategy is due to \cite{TC92} for $d>2$.
If we follow a node-selection strategy, all labels in $L_u$ will be pushed along the out-arcs of a selected node $u$.
This strategy was first proposed in \cite{BS89} for the biobjective case.
Nodes or labels which are ready to be selected are called \emph{open}.

After pushing a set of labels, the label sets at the head of each considered arc are \emph{cleaned}, i.e. all \emph{dominated} labels are removed from the modified label sets.
We say a label $\ell = (\mathbf{v}, u, \hat{\ell})$ dominates a label $\ell' = (\mathbf{v'}, u', \hat{\ell}', )$ if $\mathbf{v} \leq \mathbf{v'}$ and $\mathbf{v} \neq \mathbf{v'}$.

There are many ways in which a label or node can be selected.
A comparative study was conducted by Paixao and Santos \cite{PS07}.
For example, a pure FIFO strategy seems to work best in the aforementioned study.
But also other strategies are possible:
For example, we can sort the labels by their average cost, i.e., $\sum_{i\in \{1,\dots, d\}} \mathbf{v}_i / d$ and always select the smallest one.
A less expensive variant is due to \cite{BGM96, PS09}, where we decide depending on the top label $\ell = (\mathbf{v}, u, \hat{\ell})$ in a FIFO queue $Q$ where to place a new label $\ell' = (\mathbf{v'}, u', \hat{\ell}')$:
If $\mathbf{v}'$ is lexicographic smaller than $\mathbf{v}$, then it is placed in the front of $Q$, otherwise it is placed at the back of $Q$.

Both available computational studies on labeling algorithms for the multiobjective shortest path problem with more than 2 criteria suggest that label-selection strategies are far superior compared to node-selection strategies \cite{PS09, GM01}.

\subsection{Label Setting vs. Label Correcting Algorithms}

In the paper by Martins \cite{M84}, the author describes an algorithm using a label-selection strategy.
The algorithm selects the next label by choosing the lexicographic smallest label among all labels in $\mathcal{L} := \bigcup_{u\in V} L_u$.
In general, whenever we select a nondominated label $\ell = (\vv, u, \hat{\ell})$ in $\mathcal{L}$, this label represents a nondominated path from $s$ to $u$.
Labeling algorithms having the property that whenever we select a label we know that the represented path is a Pareto-optimal path, are called \emph{label setting algorithms}.
Labeling algorithms which do not have this property, and thus sometimes delete or correct a label, are called \emph{label correcting algorithms}.
On the plus side, in label setting algorithms, we never select a label which will be deleted in the process of the algorithm.
But selecting these labels is not trivial.
For example selecting a lexicographic smallest label requires a priority queue data structure, whereas the simplest label-selection strategy requires only a simple FIFO queue.

In many studies the label correcting algorithms are superior to the label setting variants. See for example \cite{GM01, PS09, RE09}.
It is not clear why this is the case.
One possible reason is, that the cost of the data structure can not make up for the advantage of not pushing too many unneeded labels.
Since the case is clear for the comparison of label setting and label correcting algorithms, we do not evaluate label setting algorithms in our study.

\section{Tree-Deletion Pruning} \label{sec:td}

\begin{figure}
\begin{center}
\includegraphics[scale=0.88]{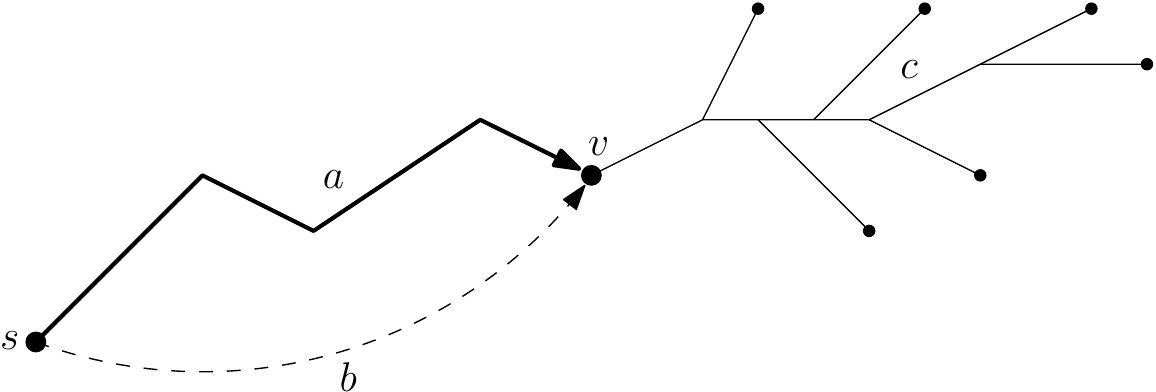}
\end{center}
\caption{Illustration of the tree-deletion pruning}
\label{fig:tree}
\end{figure}

The main difficulty which is faced by label correcting algorithms is that we can push labels which will later be dominated by a new label.
To address this issue, let us take a label-selection algorithm into consideration which selects the next label in a FIFO manner.
In Fig. \ref{fig:tree}, we see the situation where a label $\ell$ at node $v$, which encodes a path $a$ from $s$ to $v$, is dominated by a label $\ell'$, which encodes a different path $b$ from $s$ to $v$.
Based on the label $\ell$, we might already have built a tree of descendand labels $c$.
If we proceed with the usual label correcting algorithm, first the descendant labels of $\ell$ will be pushed and later be dominated by the descandants of $\ell'$.
To avoid the unnecessary pushes of descendants of $\ell$, we can delete the whole tree $c$ after the label $\ell$ is deleted.
This pruning method will be called tree-deletion pruning (TD).
We can employ this method in label correcting algorithms using both, the label-selection and node-selection strategies.

\section{Implementation Details} \label{sec:impl}

We implemented both label-correcting algorithms, a version of the FIFO label-selection (LS) and FIFO node-selection (NS) algorithms in C++11.
The reason for choosing these variants is that the LS-algorithm is the fastest method in the latest comparative study \cite{PS09}.
Both alternatives are also implemented using tree-deletion pruning (LS-TD and NS-TD, respectively).
Pseudocode can be found in Algorithms \ref{algo:ls} and \ref{algo:ns}.

\begin{algorithm}[tb]
\caption{Abstract version of the LS algorithm}
\label{algo:ls}
\begin{algorithmic}[1]
\Require Graph $G = (V, A)$, nodes $s,t\in V$ and objective function $c: A \rightarrow \QQ^d$
\Ensure List $R$ of pairs $(p,y)$ for all $y\in c(\mathcal{P}_{s,t})$ and some $p\in\mathcal{P}_{s,t}$ such that $c(p) = y$
\State $L_u \gets \emptyset$ for all $u\in V\backslash \{s\}$
\State $\ell \gets (\mathbf{0}, s, \mathbf{nil})$
\State $L_s \gets \{\ell\}$
\State $Q$.push($\ell$)
\While{not $Q$.empty}
\State $\ell = (\mathbf{v}, u, \hat{\ell})\gets$ $Q$.pop
\For{$(u, w)\in A$}
\State Push $\ell$ along $(u,w)$ and add the new label $\ell'$ to $L_w$
\State Clean $L_w$ \Comment Tree-deletion for every deleted label in $L_w$
\If{$\ell'$ is nondominated in $L_w$}
\State $Q$.push($\ell'$)
\EndIf
\EndFor
\EndWhile
\State Reconstruct paths for each label in $L_t$ and output path/vector pairs
\end{algorithmic}
\end{algorithm}

\begin{algorithm}[tb]
\caption{Abstract version of the NS algorithm}
\label{algo:ns}
\begin{algorithmic}[1]
\Require Graph $G = (V, A)$, nodes $s,t\in V$ and objective function $c: A \rightarrow \QQ^d$
\Ensure List $R$ of pairs $(p,y)$ for all $y\in c(\mathcal{P}_{s,t})$ and some $p\in\mathcal{P}_{s,t}$ such that $c(p) = y$
\State $L_u = \emptyset$ for all $u\in V\backslash \{s\}$
\State $L_s = \{(\mathbf{0}, s, \mathbf{nil})\}$
\State $Q$.push($s$)
\While{not $Q$.empty}
\State $u\gets$ $Q$.pop
\For{$(u, w)\in A$}
\For{not yet pushed label $\ell$ in $L_u$}
\State Push $\ell$ along $(u,w)$ and add the new label to $L_w$
\EndFor
\State Clean $L_w$\Comment Tree-deletion for every deleted label in $L_w$
\If{at least one new label survived the cleaning process and $w\notin Q$}
\State $Q$.push($w$)
\EndIf
\EndFor
\EndWhile
\State Reconstruct paths for each label in $L_t$ and output path/vector pairs
\end{algorithmic}
\end{algorithm}

We use the OGDF\footnote{http://ogdf.net/} for the representation of graphs.
Wherever possible, we try to use \texttt{std::vector} for collections of data.

\paragraph{Label Selection}
In the label-selection variant we use a \texttt{std::deque} to implement the queue of open labels.

\paragraph{Node Selection}
In the node-selection variants we use our own implementation of a ring-buffer to implement the queue of open nodes.
Also, only those labels of a node are pushed, which have not been pushed before.

\paragraph{Tree-Deletion Pruning}
The successors of each label are stored in an \texttt{std::list}.
The reason for this is that the sucessor-lists are constructed empty and most of them remain empty for the whole process of the algorithm.
Construction of empty \texttt{std::list}s is the cheapest operation among the creation of all other relevant data structures.

\paragraph{Cleaning Step}
While there is a considerable literature on finding the subset of minimal vectors in a set of vectors, it is not clear which method to use in practice.
In the node-selection algorithm, we could use the fact that we try to find the nondominated vectors of two sets of nondominated vectors, which has been successfully exploited for the biobjective problem in \cite{BS89}.
To make the comparison between label-selection and node-selection strategies more focused on the strategies itself, we use a simple pairwaise comparison between all pushed labels and all labels at the head of the arc.
A more sophisticated method would make the node-selection strategy only faster.
Moreover, the studies on multiobjective labeling algorithms also employ this method.

\section{Computational Study} \label{sec:study}

The experiments were performed on an Intel Core i7-3770, 3.4 GHz and 16 GB of memory running Ubuntu Linux 12.04.
We compiled the code using LLVM 3.4 with compiler flag -O3.

In the computational study, we are concerned with the following questions:
\begin{enumerate}
	\item Is label selection really faster than node selection?
	\item In which circumstances is tree-deletion pruning useful?
\end{enumerate}
We will answer these questions in the following subsections.

\subsection{Instances}

\begin{figure}[tb]
\centering
\begin{tabular}{l|c|c||l|c|c}
Name & $n$ & $d$ & Name & $n$ & $d$\\
\hline
\texttt{CompleteN-medium} & 10--200 & 3 & \texttt{CompleteN-large}&10--140&6\\
\texttt{CompleteK-medium} & 50 & 2--15 & \texttt{CompleteK-large}&100&2--9\\
\texttt{GridN-medium} & 441--1225 & 3 & \texttt{GridN-large} & 25--289& 6 \\
\texttt{GridK-medium} & 81 & 2--15 & \texttt{GridK-large} &100&2--9\\
\texttt{RandomN-medium} & 500--10000 & 3 & \texttt{RandomN-large} &1000--20000&6\\
\texttt{RandomK-medium} & 2500 & 2--15 & \texttt{RandomK-large} &5000&2--9\\
\end{tabular}
\caption{Overview on the artificial test instances}
\label{tab:instances}
\end{figure}

We used two sets of instances.
First, we took the instances of \cite{PS09} and tested our implementations on them.
The aforementioned study is the latest study which tested implementations of labelling algorithms for MOSP.
For each problem type, there are 50 randomly drawn instances.
In summary they make up a set of 9,050 test problems.

The properties of these instances can be seen in Fig. \ref{tab:instances}.
The random graph instances are based on a hamiltonian cycle where arcs are randomly added to the graph.
In the complete graph, arcs are added between each pair of nodes in both directions.
The grid graphs are all square.
The arc costs were choosen uniformly random in $[1, 1000] \cap \NN$.
The instances are available online, see \cite{PS09} for details.

The second set of instances is similar to the instances from \cite{DW09}.
They are based on the road network of Western Europe provided by PTV AG for scientific use.
We conducted our experiments on the road network of the Czech Republic (CZE, 23,094 nodes, 53,412 edges), Luxembourg (LUX, 30,661 nodes, 71,619 edges), Ireland (IRL, 32,868 nodes, 71,655 edges) and Portugal (PRT, 159,945 nodes, 372,129 edges, only for the tree-deletion experiments).
As metrics we used similar metrics as in \cite{DW09}: travel distance, cost based on fuel consumption and travel time.
The median Pareto-front sizes are 13.0, 30.5, 12.5 and 133.5, respectively and thus comparable or larger than those in \cite{DW09}.
For each of the instances we drew 50 pairs of source and target nodes.

\subsection{Running Times Node Selection vs. Label Selection}

\begin{figure}[tb]
\centering
\subfloat[\texttt{RandomK-large}]{
	\includegraphics[scale=0.42]{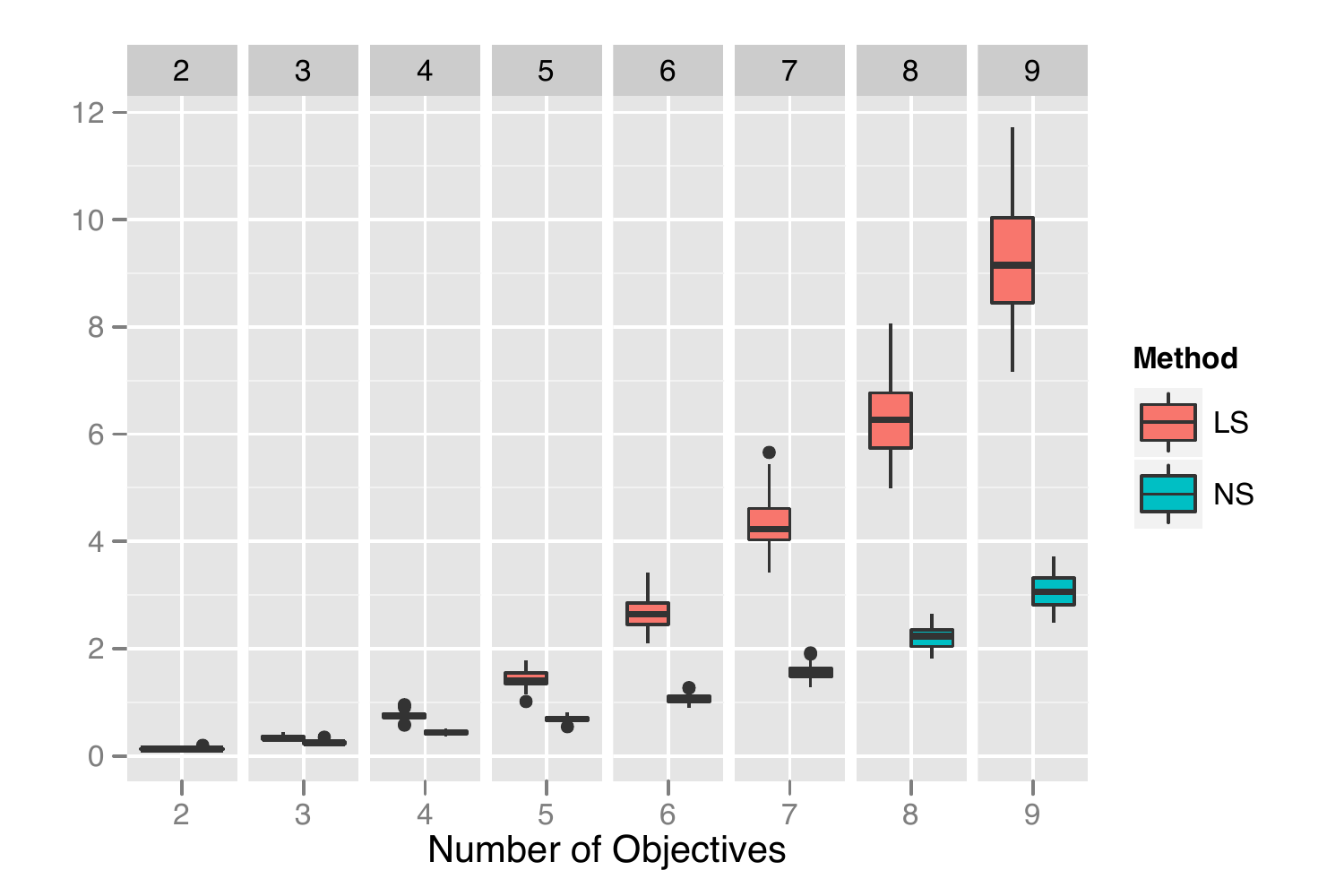}
}
\subfloat[Real-world road networks]{
	\includegraphics[scale=0.42]{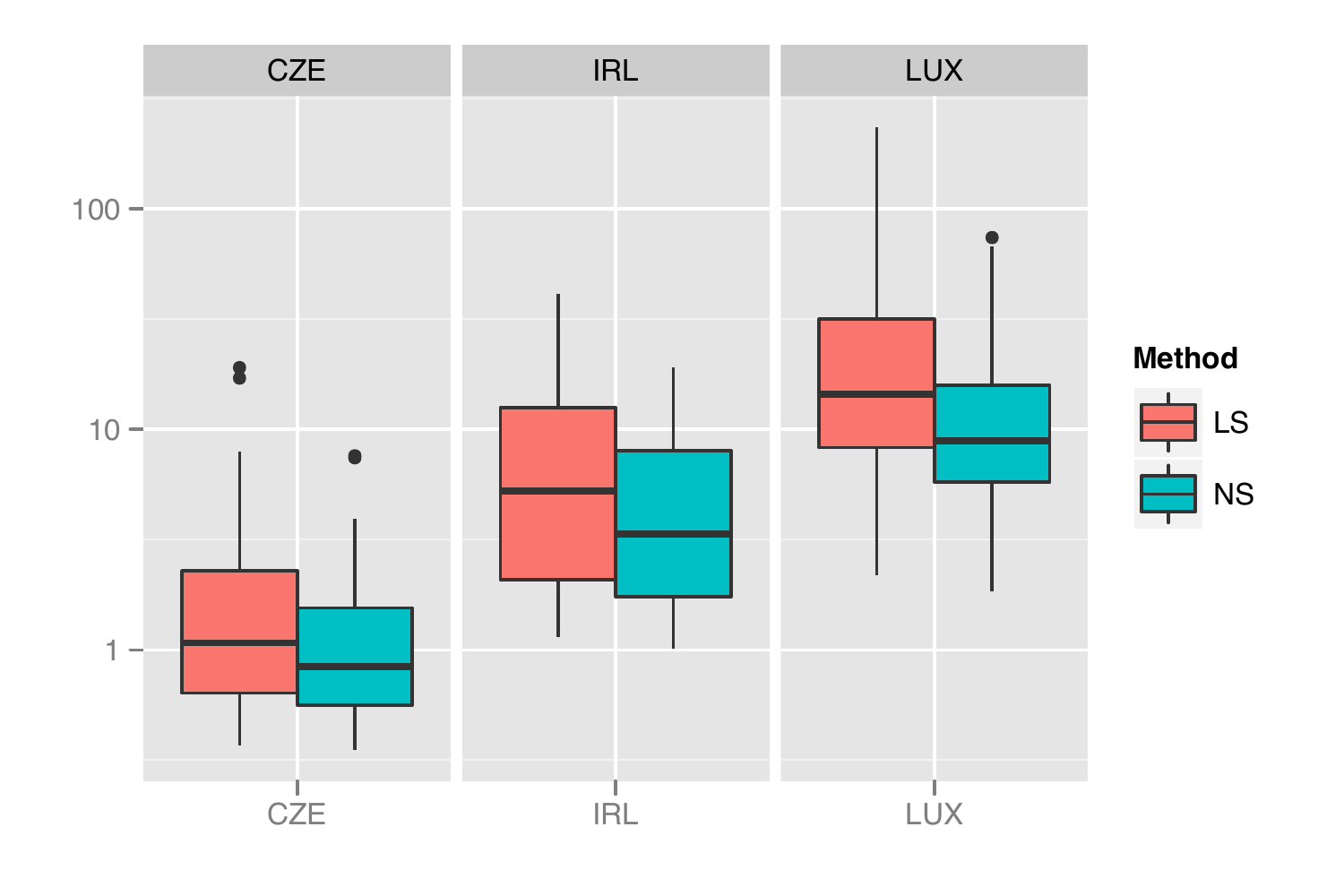}
}\\
\subfloat[\texttt{RandomN-large}]{
	\includegraphics[scale=0.42]{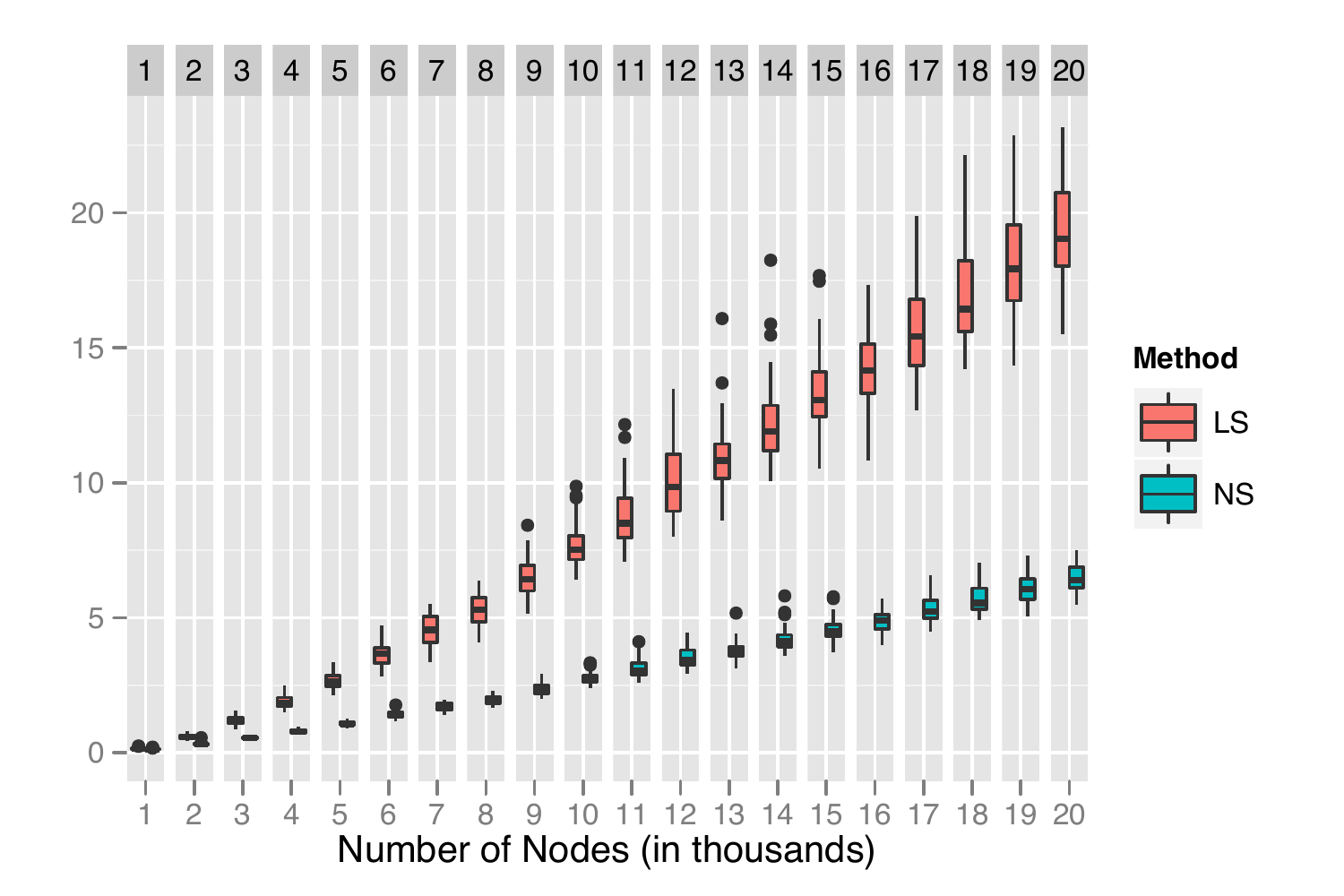}
}
\subfloat[\texttt{GridK-large}]{
	\includegraphics[scale=0.42]{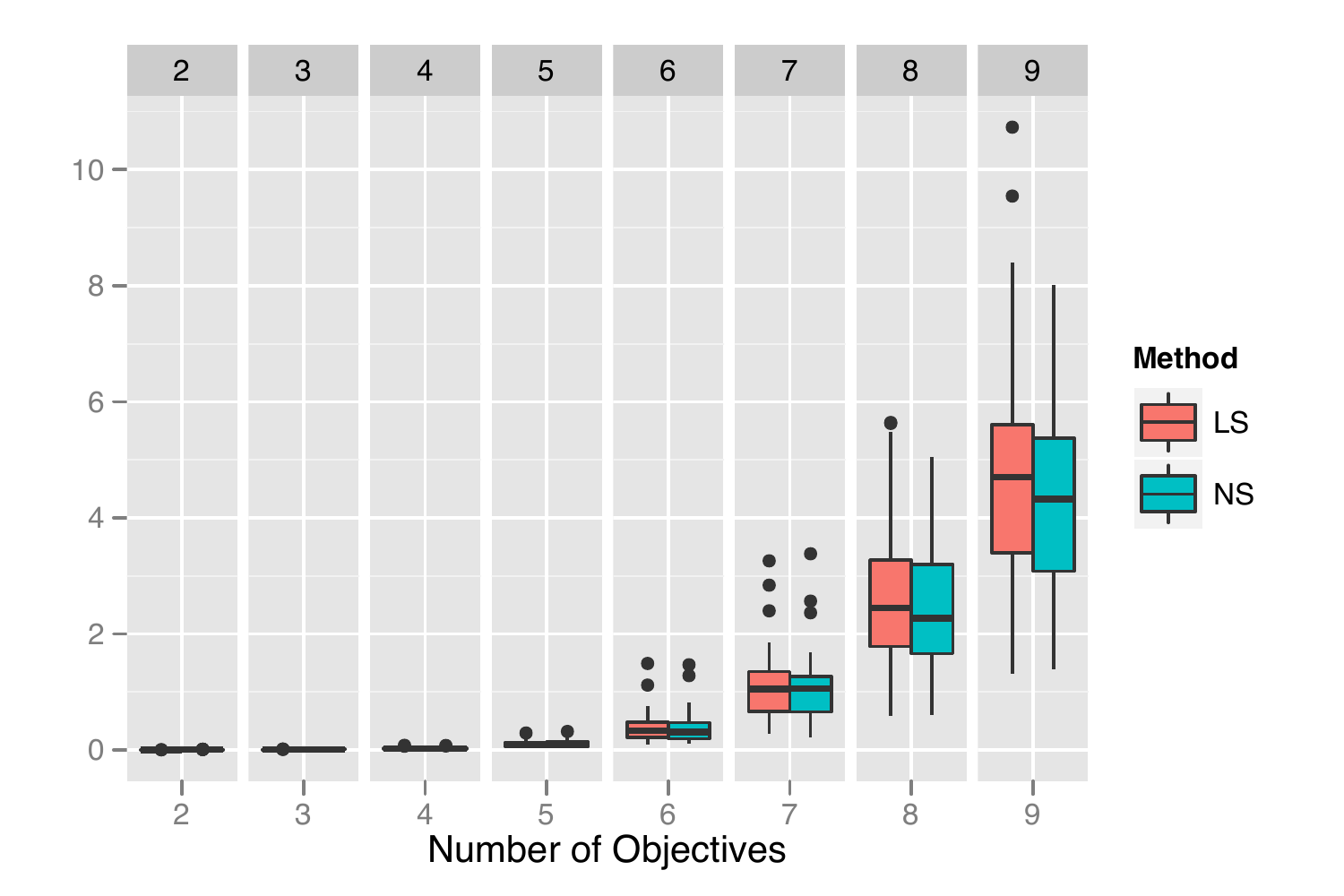}
}
\caption{Comparison of the running times (in seconds) of the label-selection (LS) and node-selection (NS) strategies}
\label{fig:ls-ns}
\end{figure}

In Fig. \ref{fig:ls-ns} we see a selection of the results on the \texttt{RandomN-large}, \texttt{RandomK-large}, \texttt{GridK-large} as well as the real-world road network instance sets.

To evaluate the results we decided to show box plots, because the deviation of the running times is very large and it is easier to recognize trends.
The box plots give a direct overview on the quartiles (box dimension), median (horizontal line inside the box), deviation from the mean (\emph{whiskers}: lines above and below the box) and outliers (points above and below the whiskers).

We see that the node-selection strategy performs better on all these instances than the label-selection strategy.
The node-selection strategy is up to a factor of 3 faster on both test sets.
This is also true for the other large and medium sized instances where the maximum factors range from 1.4 to 3.17.
Detailed box plots can be found in the appendix.

Also on the real-world road networks the results are positive.
The node-selection strategy is up to factor of 3.16 faster than the label-selection strategy.

A partial explanation can be attributed to memory management:
In the node-selection strategy a consecutive chunk of memory which contains the values of the labels pushed along an arc can be accessed in one cache access.
While in the label-selection strategy only one label is picked in each iteration, producing potentially many cache misses when the next label---potentially at a very different location---is accessed.

\subsection{Tree-Deletion Pruning}

\begin{figure}[tb]
\centering
\subfloat[\texttt{CompleteN-large}]{
	\includegraphics[scale=0.42]{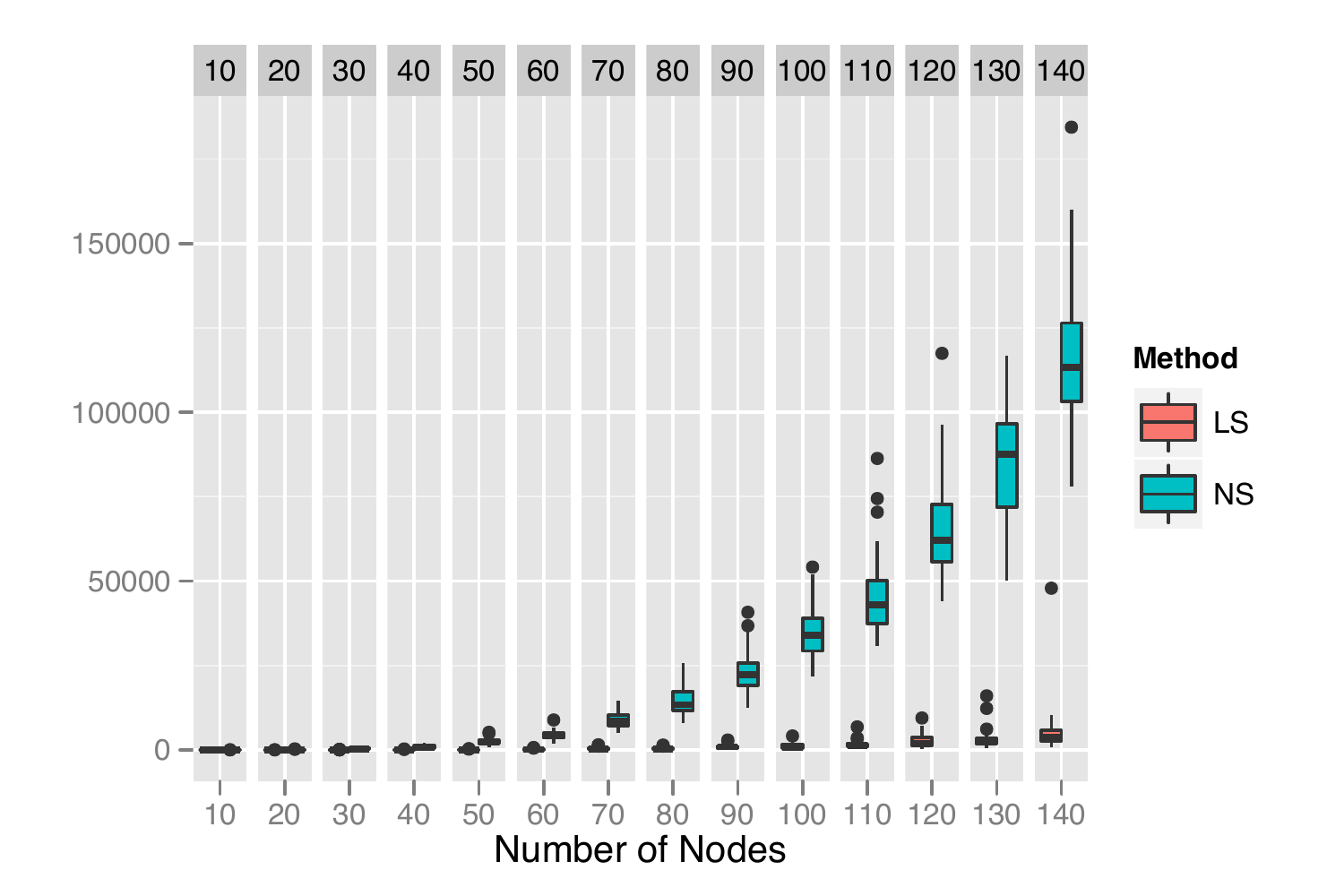}
}
\subfloat[\texttt{CompleteK-large}]{
	\includegraphics[scale=0.42]{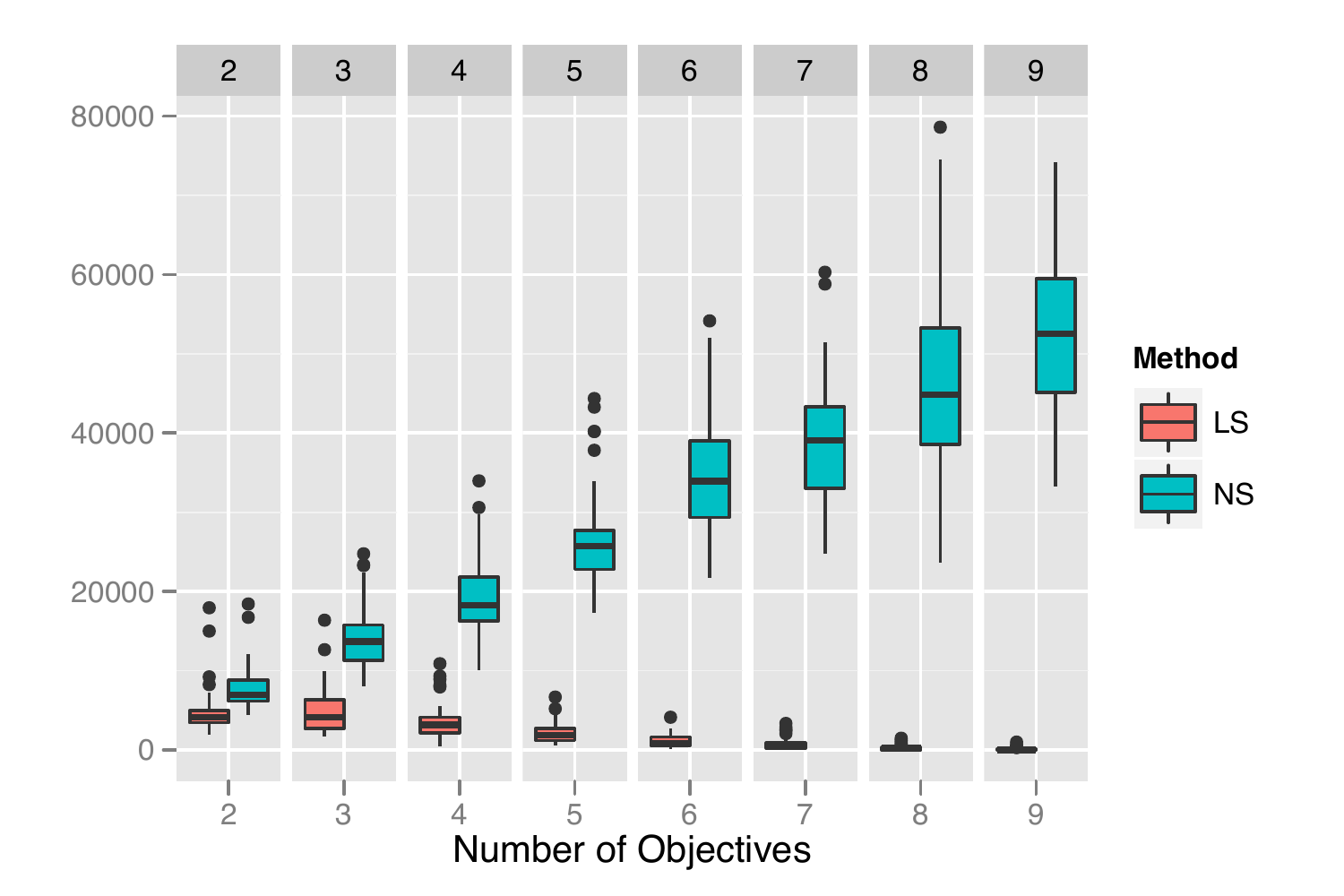}
}\\
\subfloat[\texttt{GridK-large}]{
	\includegraphics[scale=0.42]{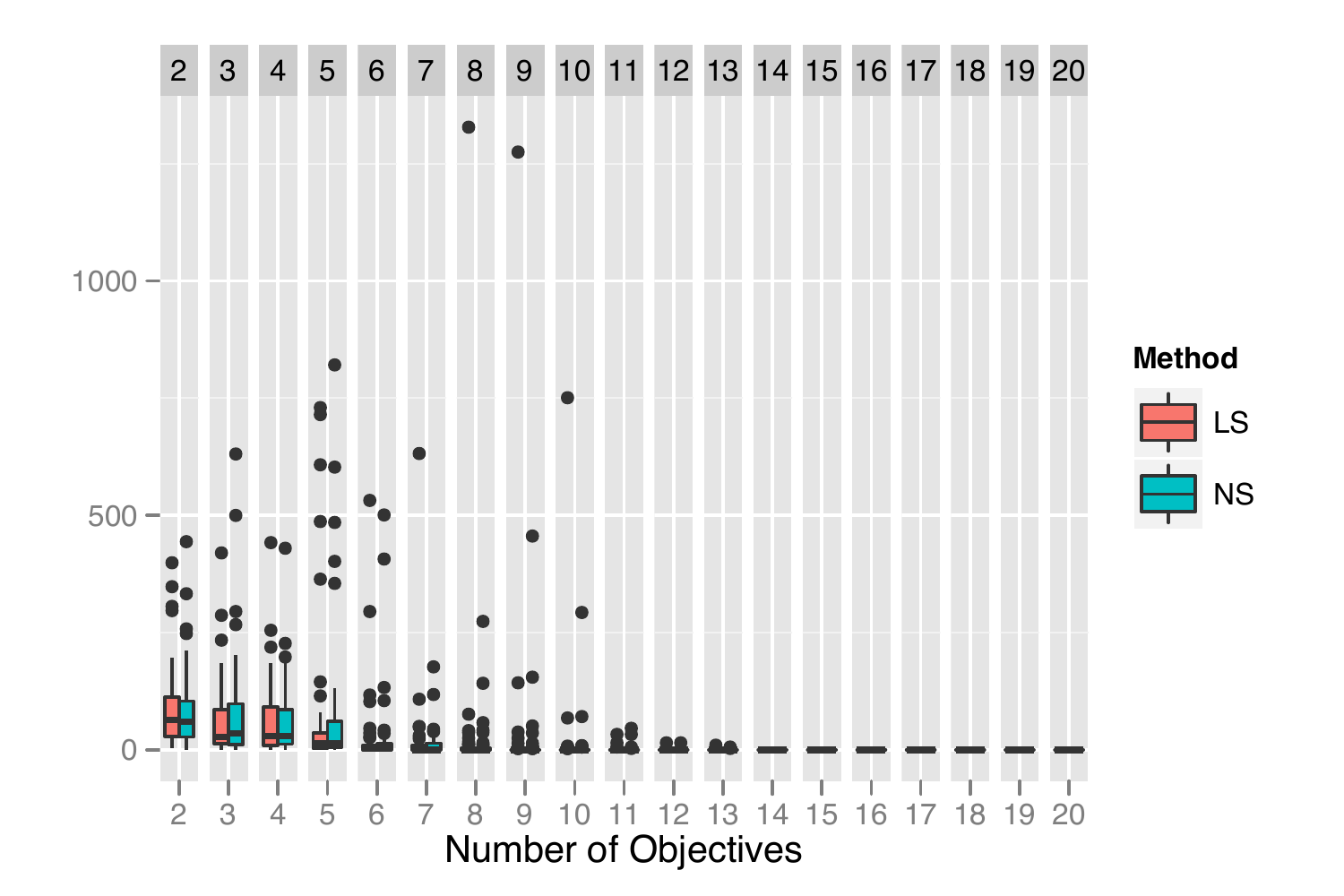}
}
\subfloat[\texttt{RandomK-large}]{\label{fig:rec-touch:random}
	\includegraphics[scale=0.42]{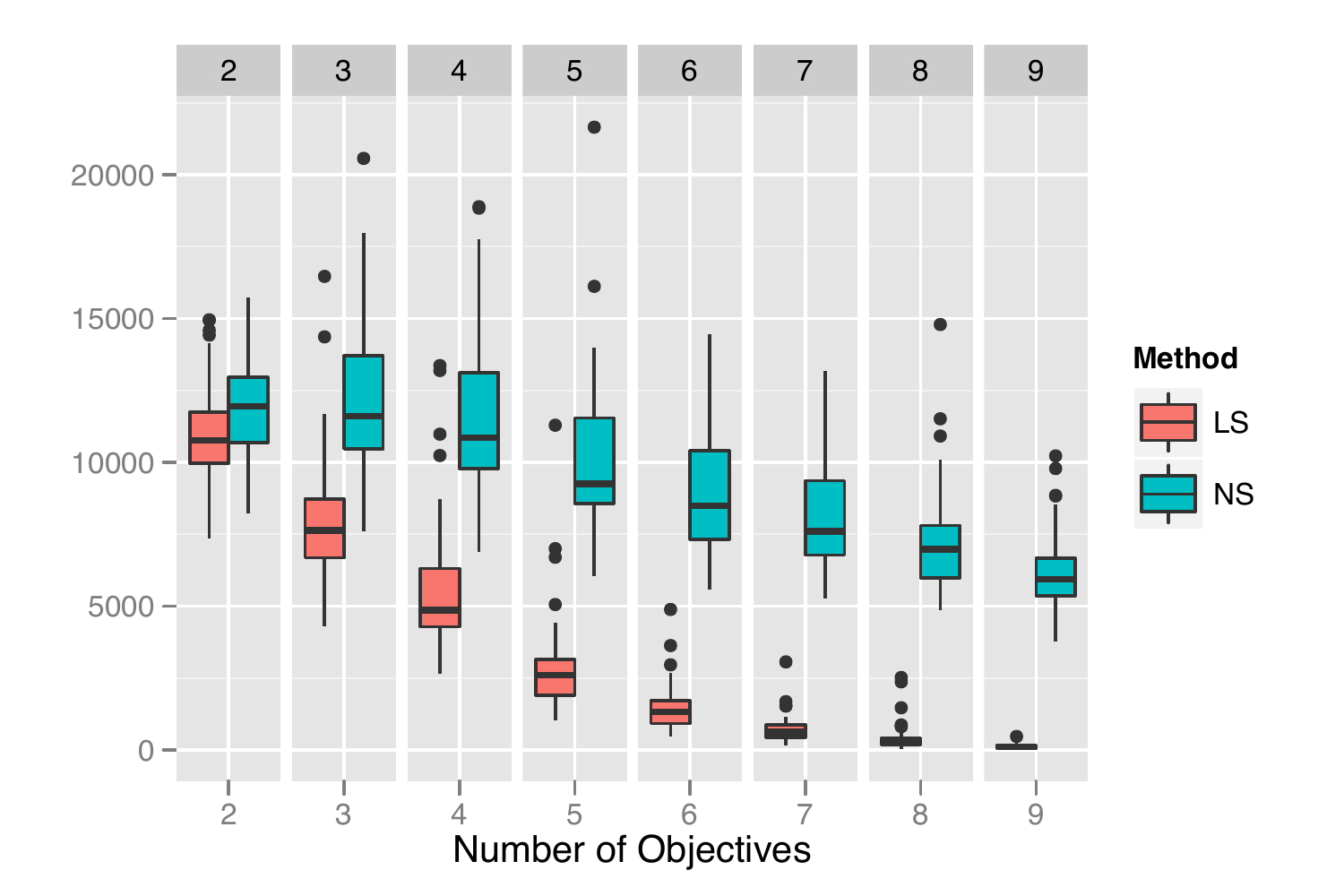}
}
\caption{Measuring how many nodes have been touched which could have been deleted by tree-deletion pruning in the label-selection (LS) and node-selection (NS) strategies}
\label{fig:rec-touch}
\end{figure}

The results concerning TD are ambiguous.
First, to see how well TD might work, we performed a set of experiments showing how many labels are touched by the label correcting algorithms which could have been deleted when using tree-deletion pruning.
In Fig. \ref{fig:rec-touch}, we see the results of this experiment on the \texttt{CompleteN-large}, \texttt{CompleteK-large}, \texttt{GridK-large} and \texttt{RandomK-large} test sets.

We observe that on these instances, especially the node-selection strategy tends to produce larger obsolete trees than the label-selection strategy.
We also observe this behavior on the complete-graph instances of medium and large size.
The situation is different on the grid-graph instances, where both algorithms have a similar tendency to produce obsolete trees.

Another observation is that when increasing the number of objectives, the number of obsolete trees which could have been deleted decreases in the grid and random graph instances (see Figs. \ref{fig:rec-touch} b and c).
This happens because when looking at instances with a large number of objectives and totally random objective values, most labels remain nondominated in the cleaning step.
The complete-graph instances are an exception here, because the number and kind of labels pushed to each node are very diverse und so still many labels are dominated.

\begin{figure}[tb]
\centering
\subfloat[\texttt{GridK-medium}]{
	\includegraphics[scale=0.42]{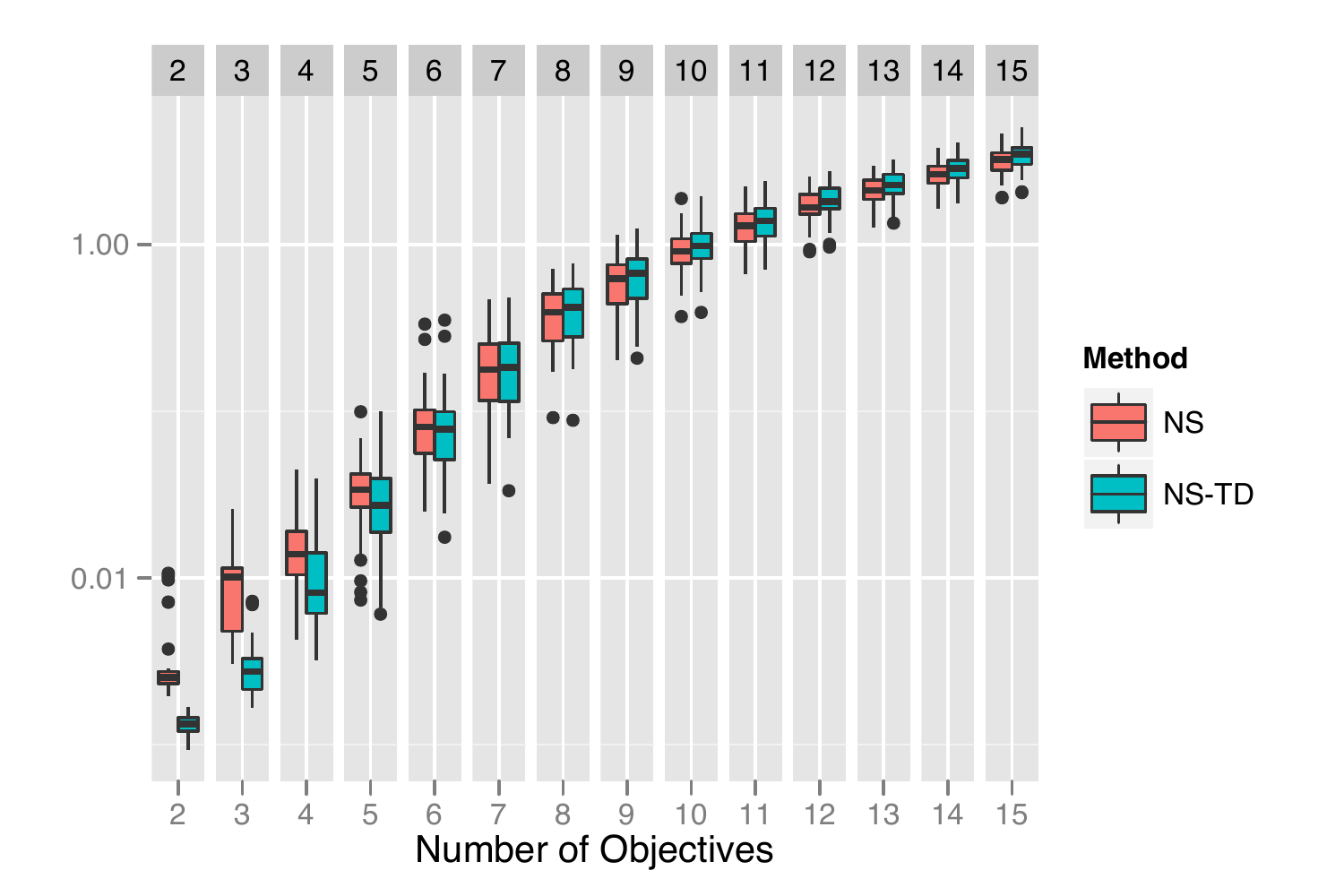}
}
\subfloat[\texttt{Real-world road networks}]{
	\includegraphics[scale=0.42]{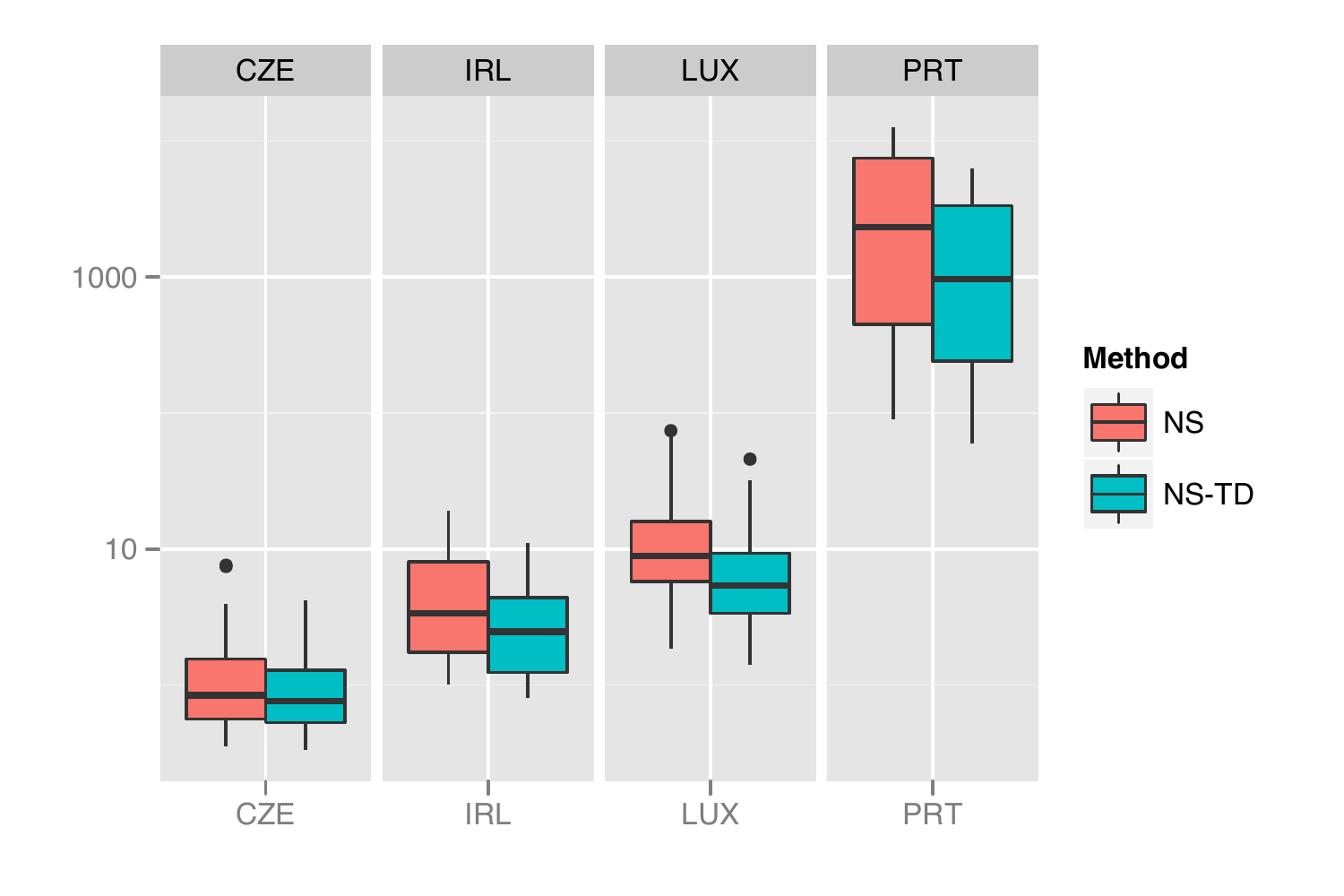}
}\\
\subfloat[\texttt{RandomN-large}]{
	\includegraphics[scale=0.42]{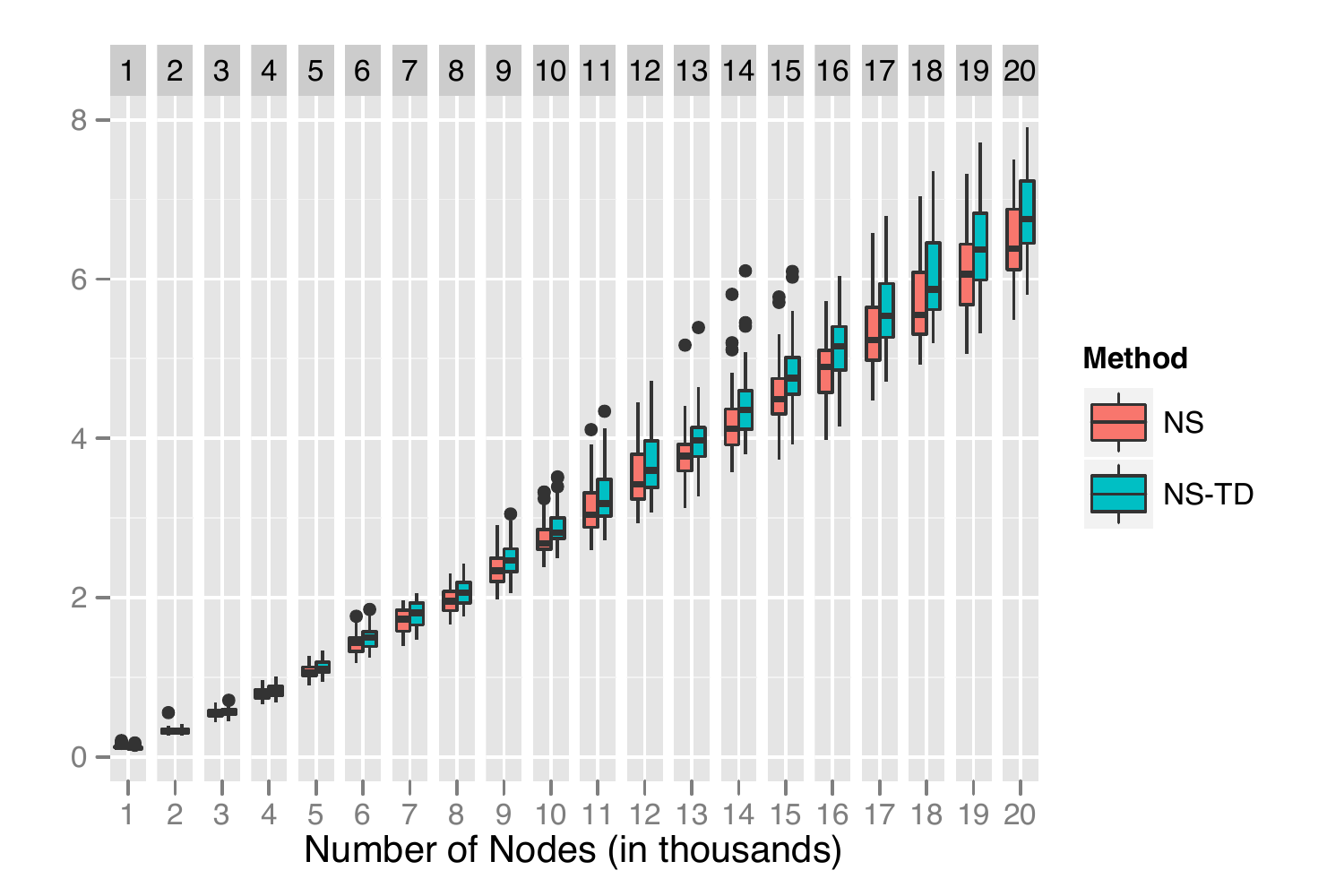}
}
\subfloat[\texttt{CompleteN-large}]{
	\includegraphics[scale=0.42]{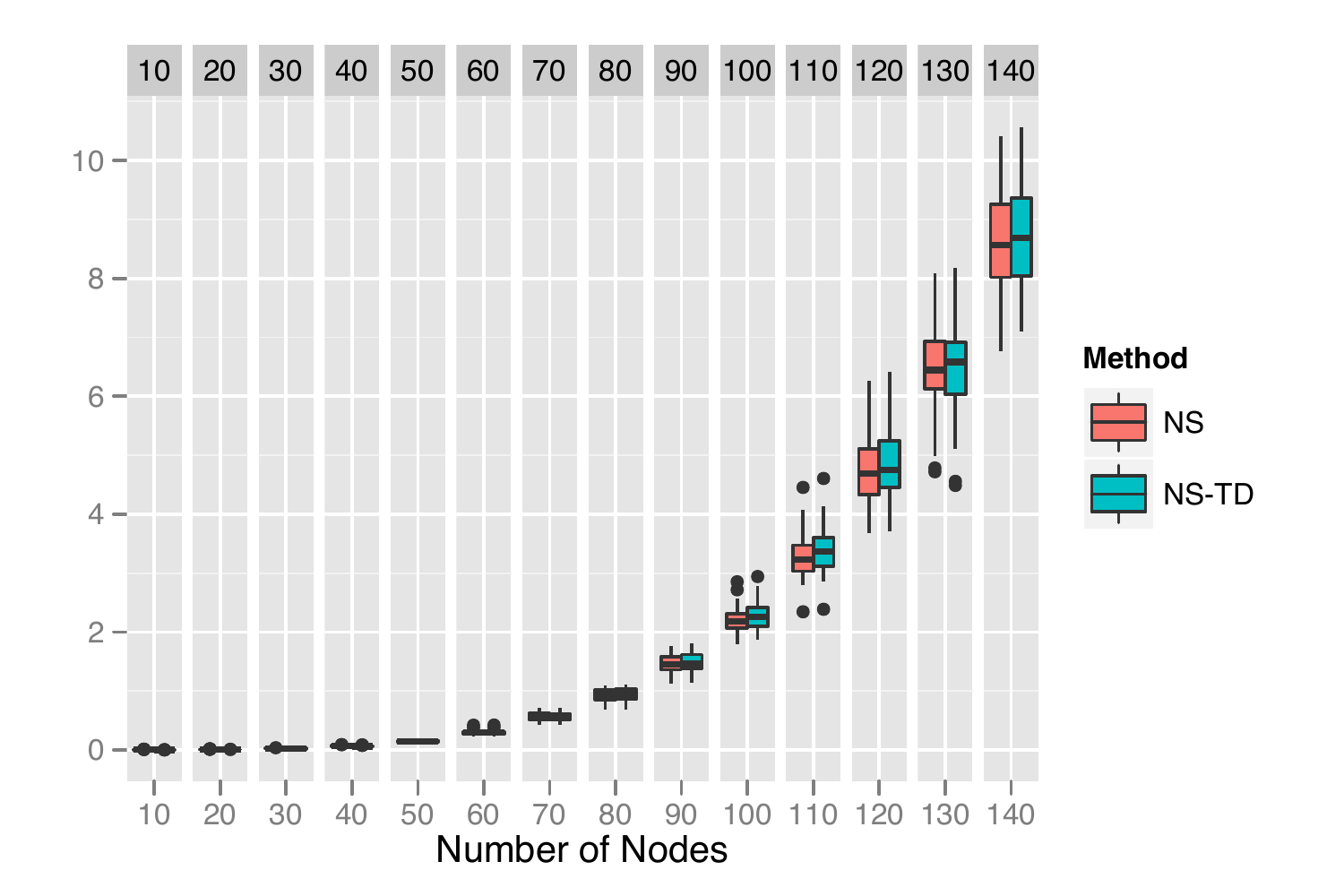}
}
\caption{Comparison of the running times (in seconds) of the node-selection strategy with (NS-TD) and without (NS) tree-deletion pruning}
\label{fig:ns-td}
\end{figure}

Hence, what we expect is that on instances where a large number of labels is dominated in the cleaning step the tree-deletion pruning is very useful.
The results of the comparison of the running times can be seen in Fig. \ref{fig:ns-td}.
It can be seen on the real-world road networks that the tree-deletion pruning works very well and we can achieve a speed-up of up to 3.5 in comparison to the pure node-selection strategy.

On the artificial benchmark instances however, the results are not so clear.
TD works well on medium sized grid graphs with a small number of objectives and also on complete graphs of any size.
On the other instances of the artificial benchmark set, TD performs slightly worse than the pure node-selection strategy, especially on the \texttt{large} instances.

This behavior can be explained by the large number of labels which are dominated in the road network instances.
The size of the Pareto-fronts are small compared to the instances of the artificial test set.
So, we hypothesize that the pruning strategy is especially useful if many labels are dominated in the cleaning step and large obsolete trees can be deleted in this process.
TD seems also to work better on denser networks.

To test this hypothesis, we created a new set of random graphs.
To make the graphs more dense than in the previous instances, we drew $0.3$ times the possible number of edges and to match the small sized Pareto-fronts of the instances from \cite{DW09} with a high correlation of the objective functions, i.e., we used a Gauss copula distribution with a fixed correlation of 0.7.
If the hypothesis is false, TD should run slower than the pure node-selection strategy on these instances.

\begin{figure}[tb]
\centering
\includegraphics[scale=0.44]{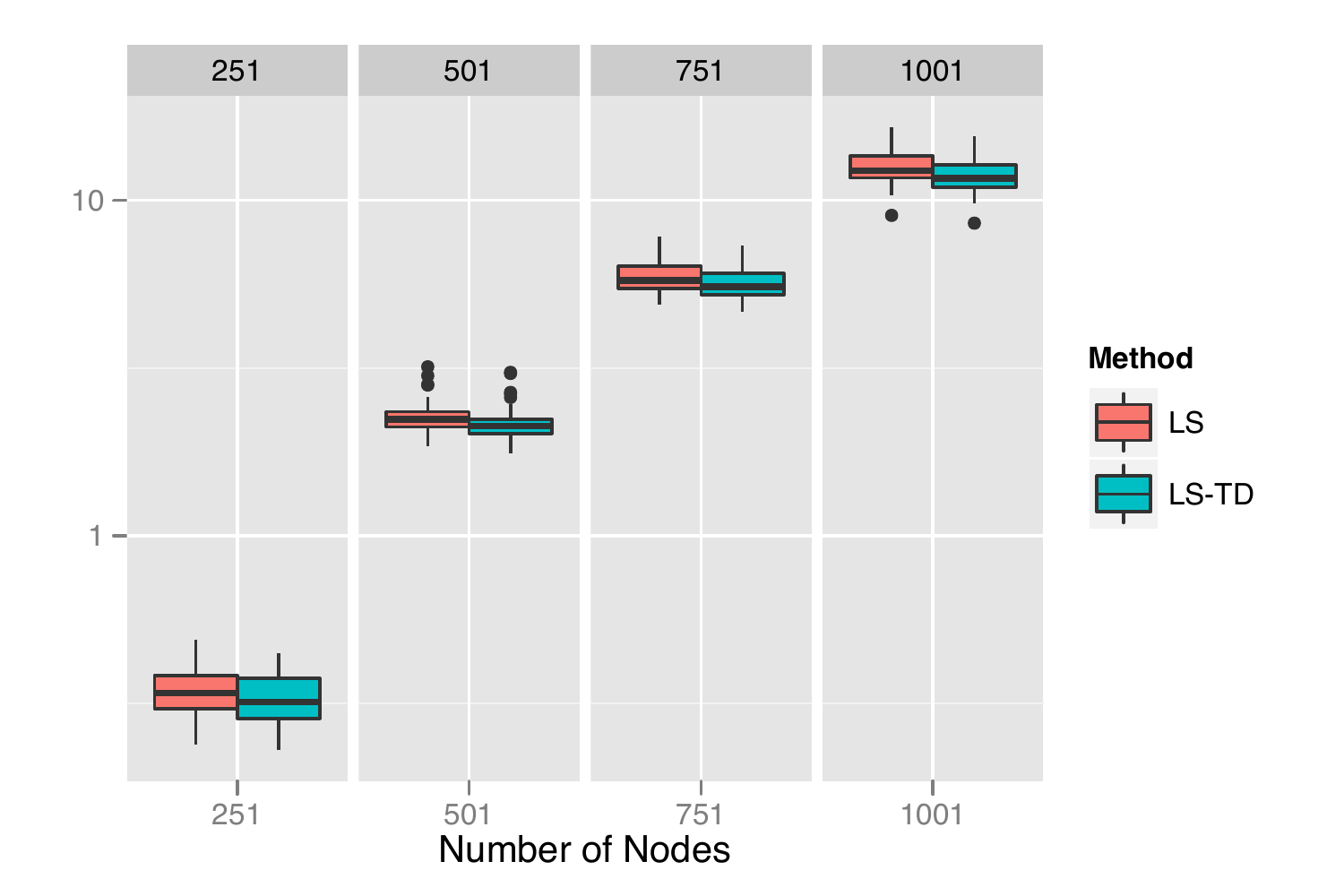}
\caption{Comparison of the running times (in seconds) of the node-selection strategy with and without TD on the corrolated random networks}
\label{fig:high_corr}
\end{figure}

But the results in Fig. \ref{fig:high_corr} show that TD beats the pure node-selection strategy on these graphs.
TD achieves a speed-up of up to 1.14.
Using a wilcoxon signed rank test we can also see that the hypothesis that TD is slower than the pure node-selection strategy on these instances can be refused with a p-value of less than 0.001.

\section{Conclusion}

To conclude, we showed in this paper that node-selection strategies in labeling algorithms for the MOSP problem can be advantageous, especially if implemented carefully.
So node-selection strategies should not be neglected as an option for certain instance classes.

We showed that the tree-deletion pruning we introduced in this paper, works well on the real-world road networks.
On the artificial instances it does not seem to work to well, which we can explain by the very low densities and unrealistic objective functions used in these instances.
To show that TD works well when having larger correlations as in the real-world road networks and higher densities, we also created instances which had the potential to refute this hypothesis.
But the hypothesis passed the test.


\bibliographystyle{splncs_srt}
\bibliography{bibliography}

\FloatBarrier
\newpage
\appendix

\section*{Appendix}

\begin{figure}[tb]
\centering
\makebox[0cm]{
	\subfloat[\texttt{CompleteN-medium}]{
		\includegraphics[scale=0.51]{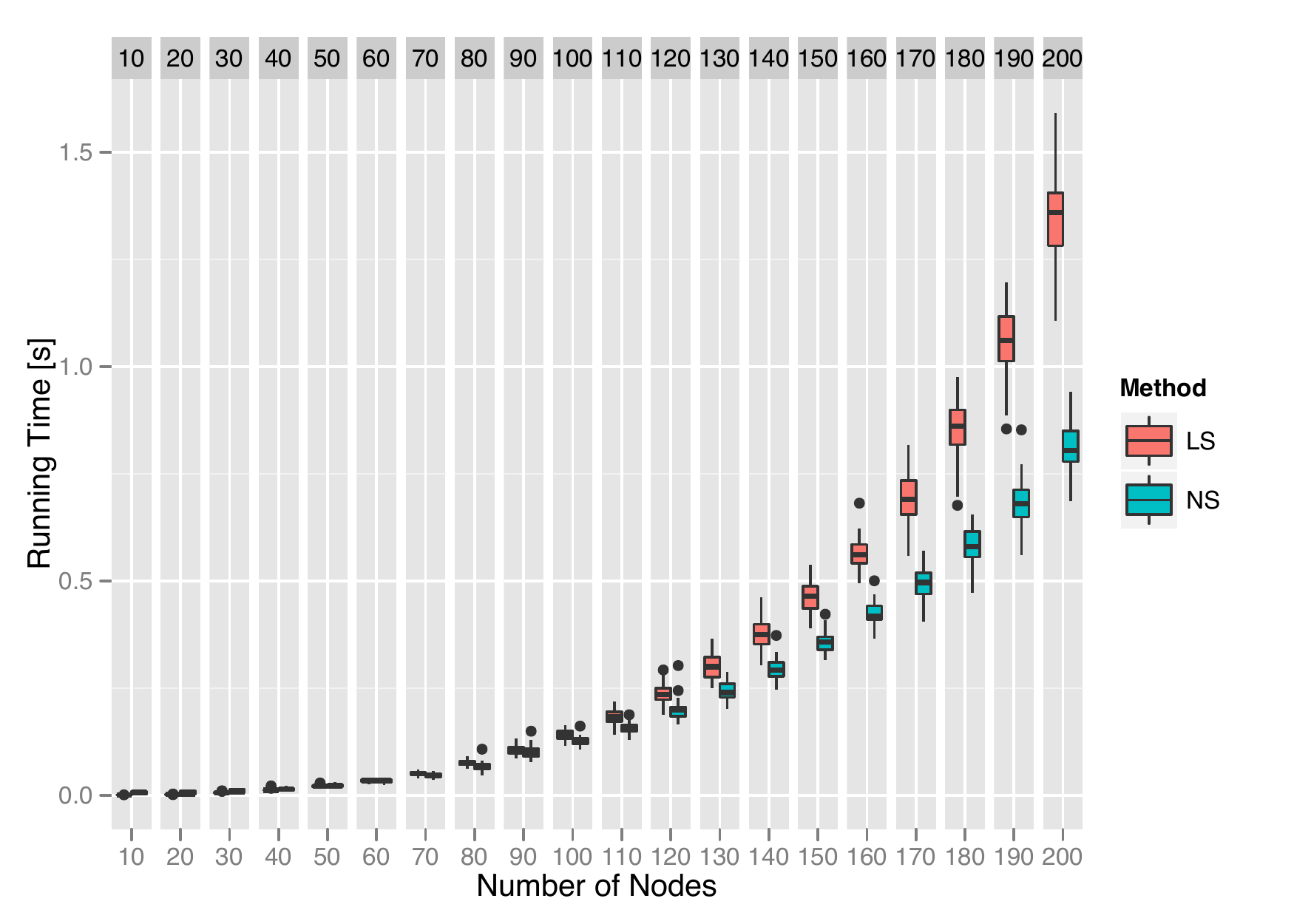}
	}
	\subfloat[\texttt{CompleteK-medium}]{
		\includegraphics[scale=0.51]{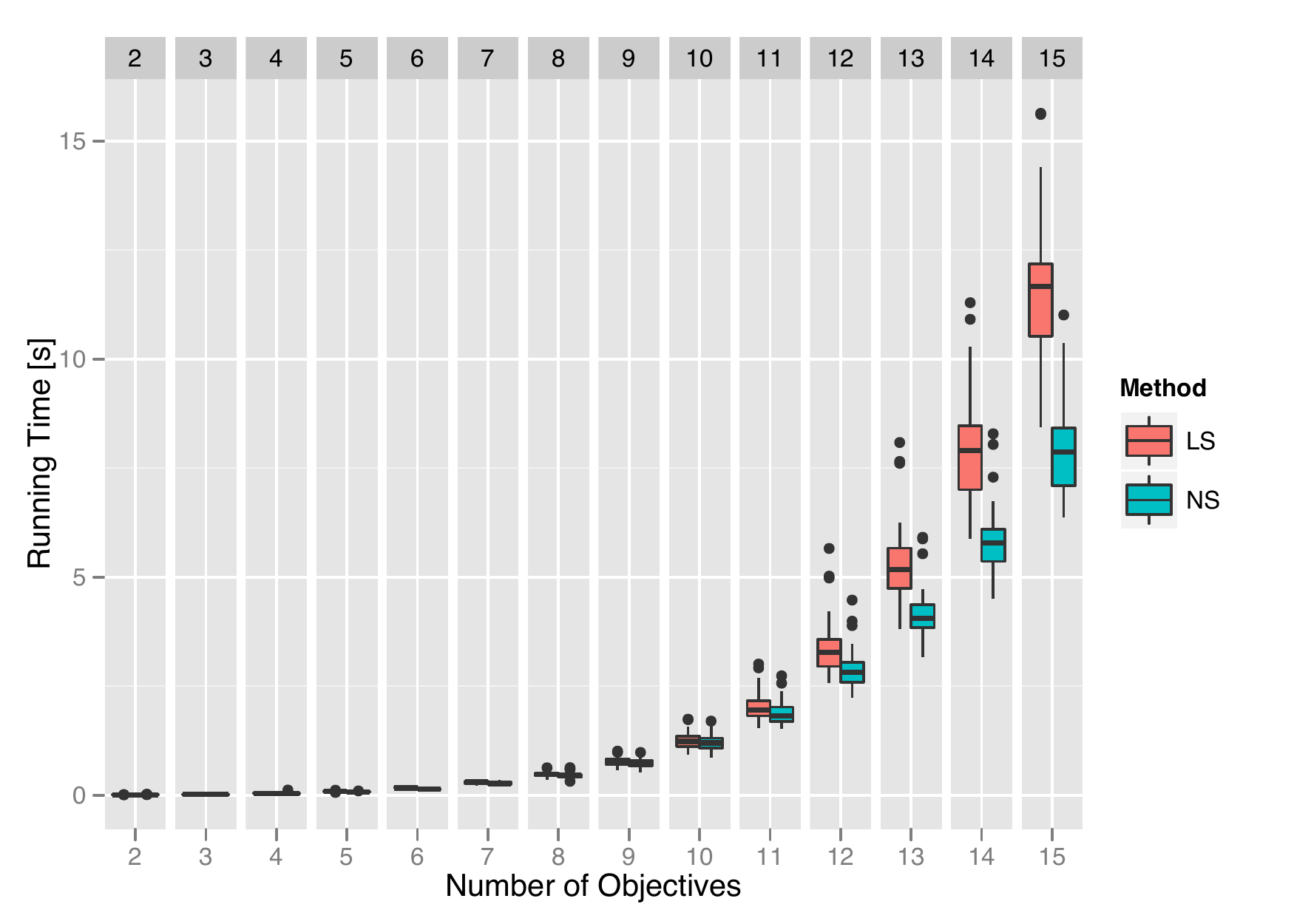}
	}
}\\
\makebox[0cm]{
	\subfloat[\texttt{GridN-medium}]{
		\includegraphics[scale=0.51]{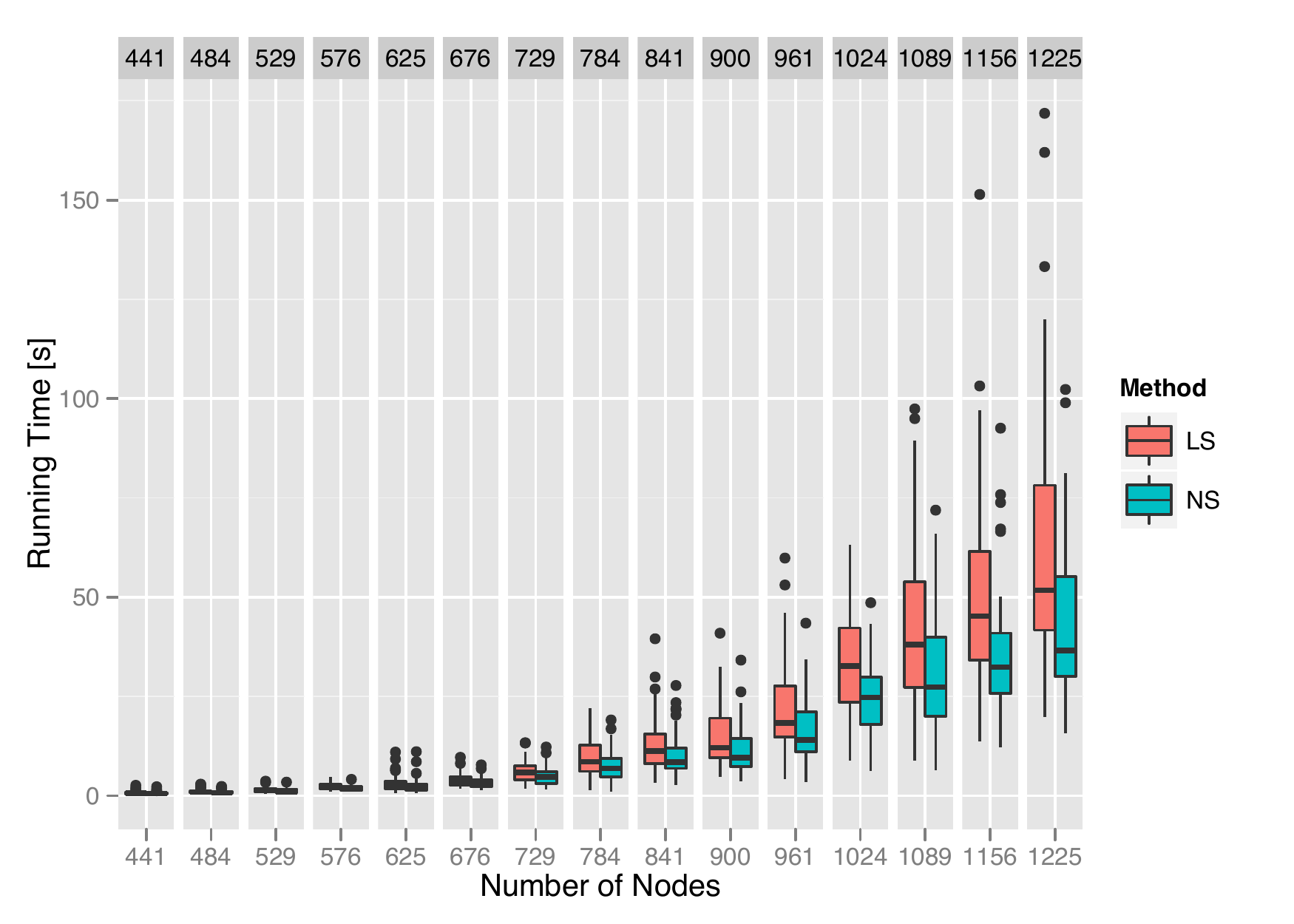}
	}
	\subfloat[\texttt{GridK-medium}]{
		\includegraphics[scale=0.51]{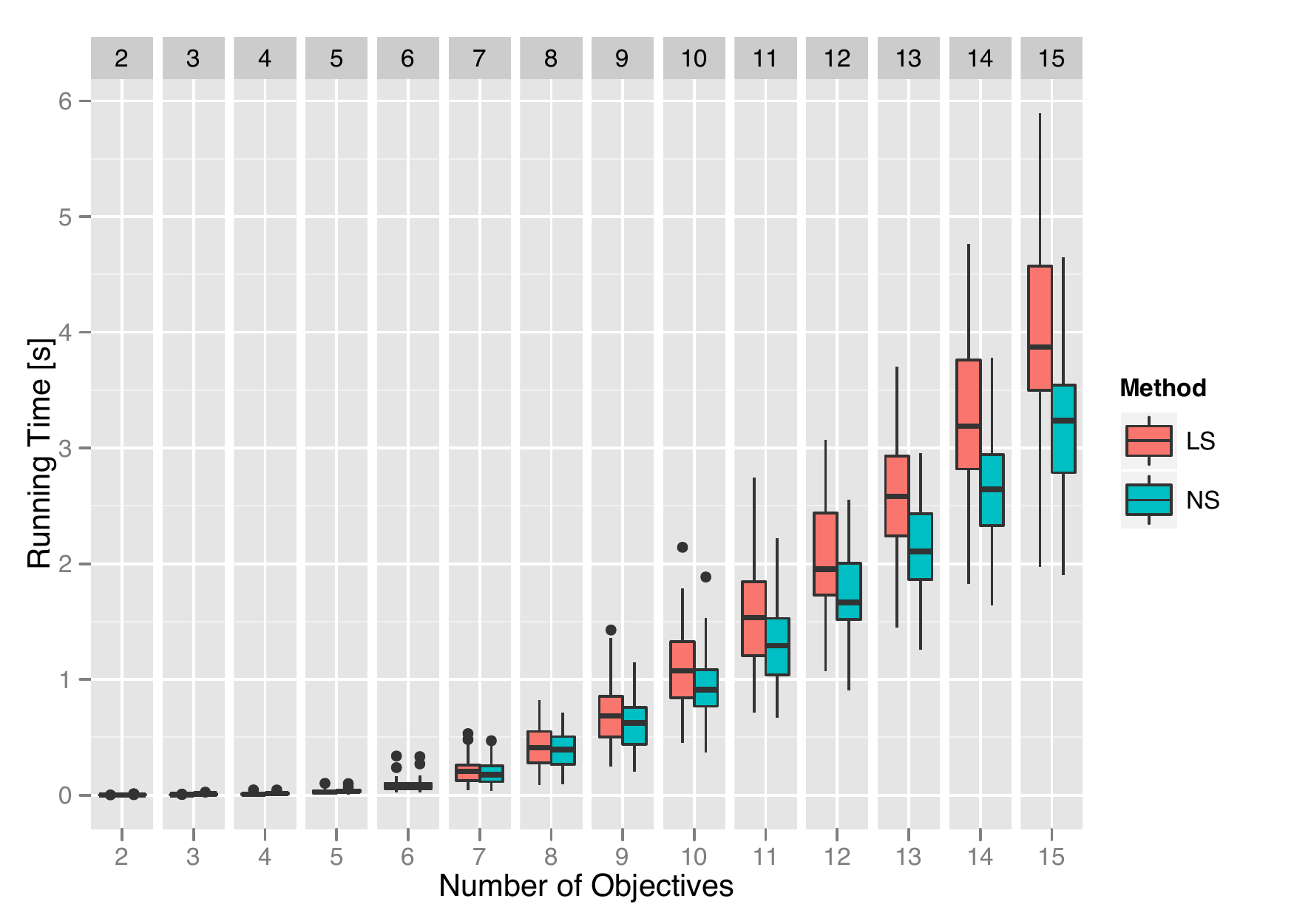}
	}
}\\
\makebox[0cm]{
	\subfloat[\texttt{RandomN-medium}]{
		\includegraphics[scale=0.51]{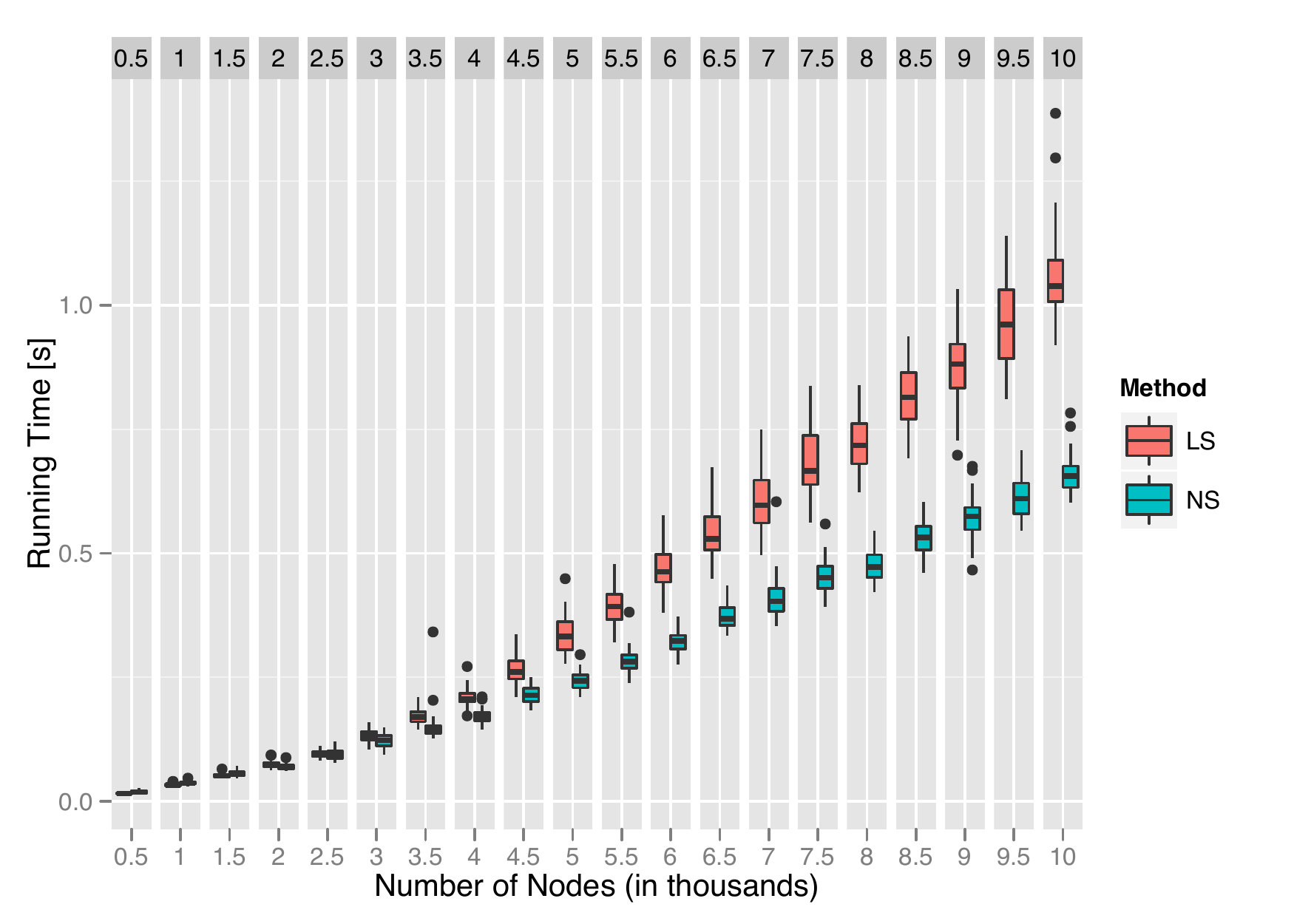}
	}
	\subfloat[\texttt{RandomK-medium}]{
		\includegraphics[scale=0.51]{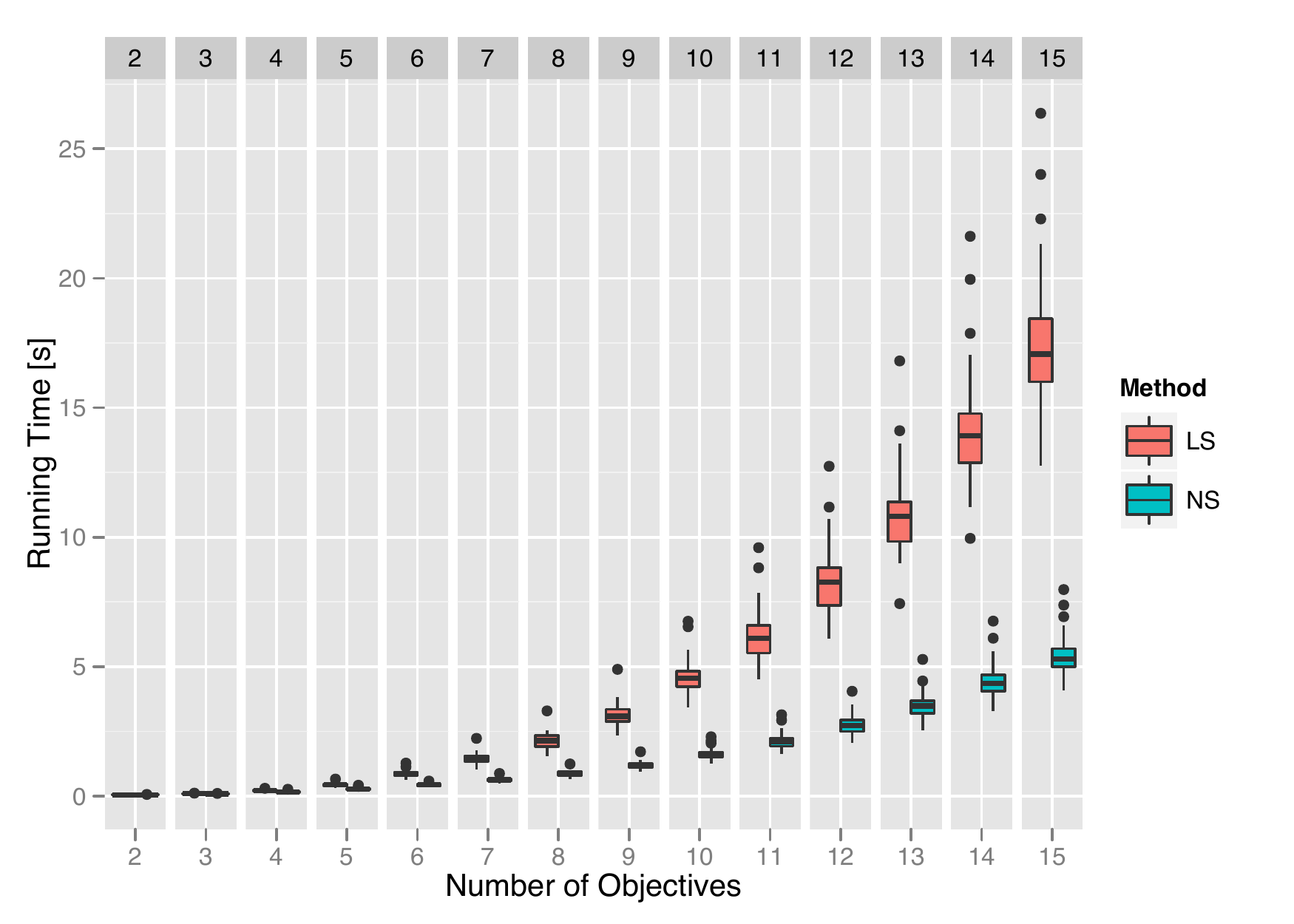}
	}
}
\caption{Comparison of the running times of the label-selection (LS) and node-selection (NS) strategies}
\label{fig:ls-ns:medium}
\end{figure}

\begin{figure}[tb]
\centering
\makebox[0cm]{
	\subfloat[\texttt{CompleteN-large}]{
		\includegraphics[scale=0.51]{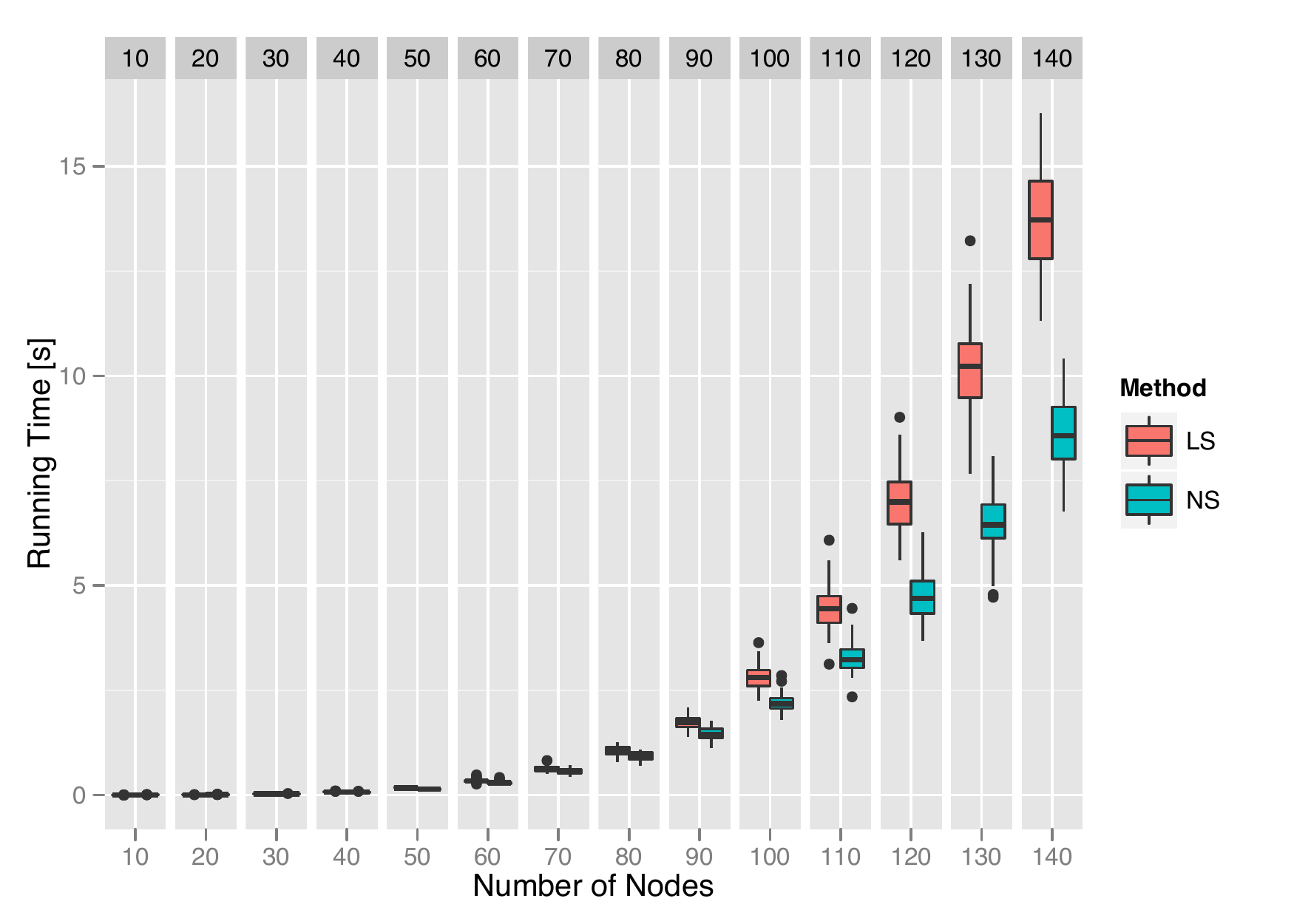}
	}
	\subfloat[\texttt{CompleteK-large}]{
		\includegraphics[scale=0.51]{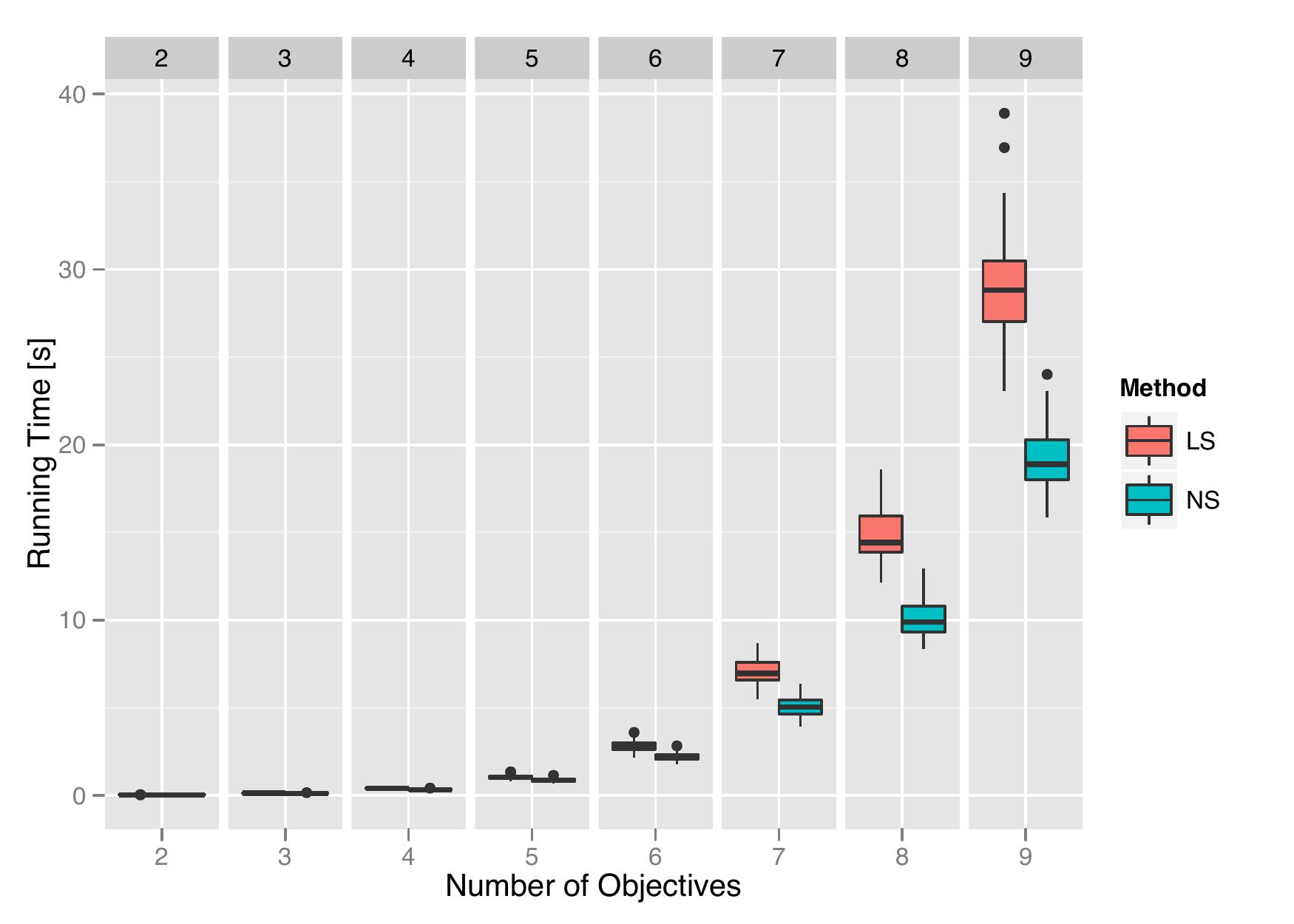}
	}
}\\
\makebox[0cm]{
	\subfloat[\texttt{GridN-large}]{
		\includegraphics[scale=0.51]{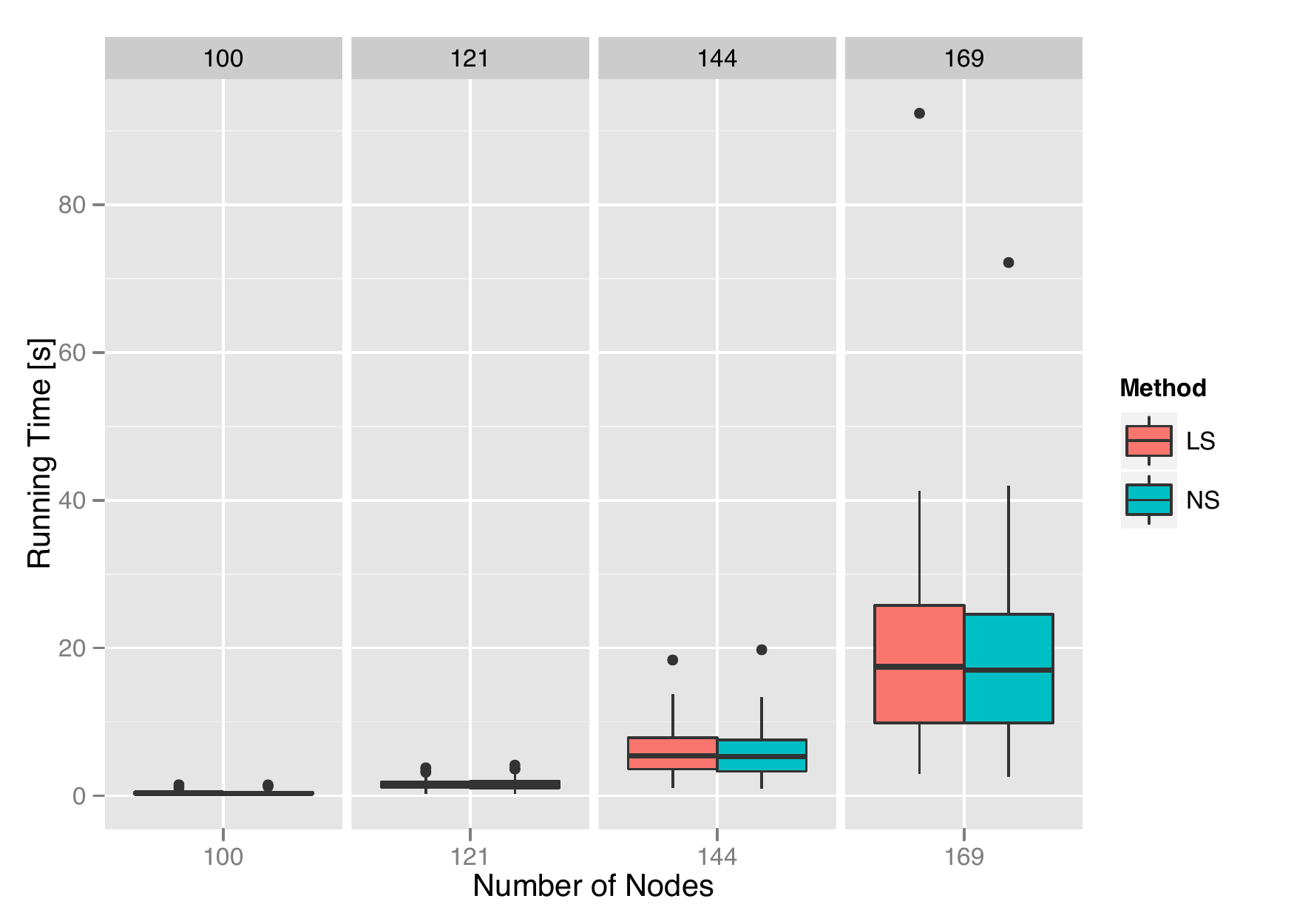}
	}
	\subfloat[\texttt{GridK-large}]{
		\includegraphics[scale=0.6]{ls-ns_GridK-large}
	}
}\\
\makebox[0cm]{
	\subfloat[\texttt{RandomN-large}]{
		\includegraphics[scale=0.6]{ls-ns_RandomN-large}
	}
	\subfloat[\texttt{RandomK-large}]{
		\includegraphics[scale=0.6]{ls-ns_RandomK-large}
	}
}
\caption{Comparison of the running times of the label-selection (LS) and node-selection (NS) strategies}
\label{fig:ls-ns:large}
\end{figure}

\begin{figure}[tb]
\centering
\makebox[0cm]{
	\subfloat[\texttt{CompleteN-medium}]{
		\includegraphics[scale=0.51]{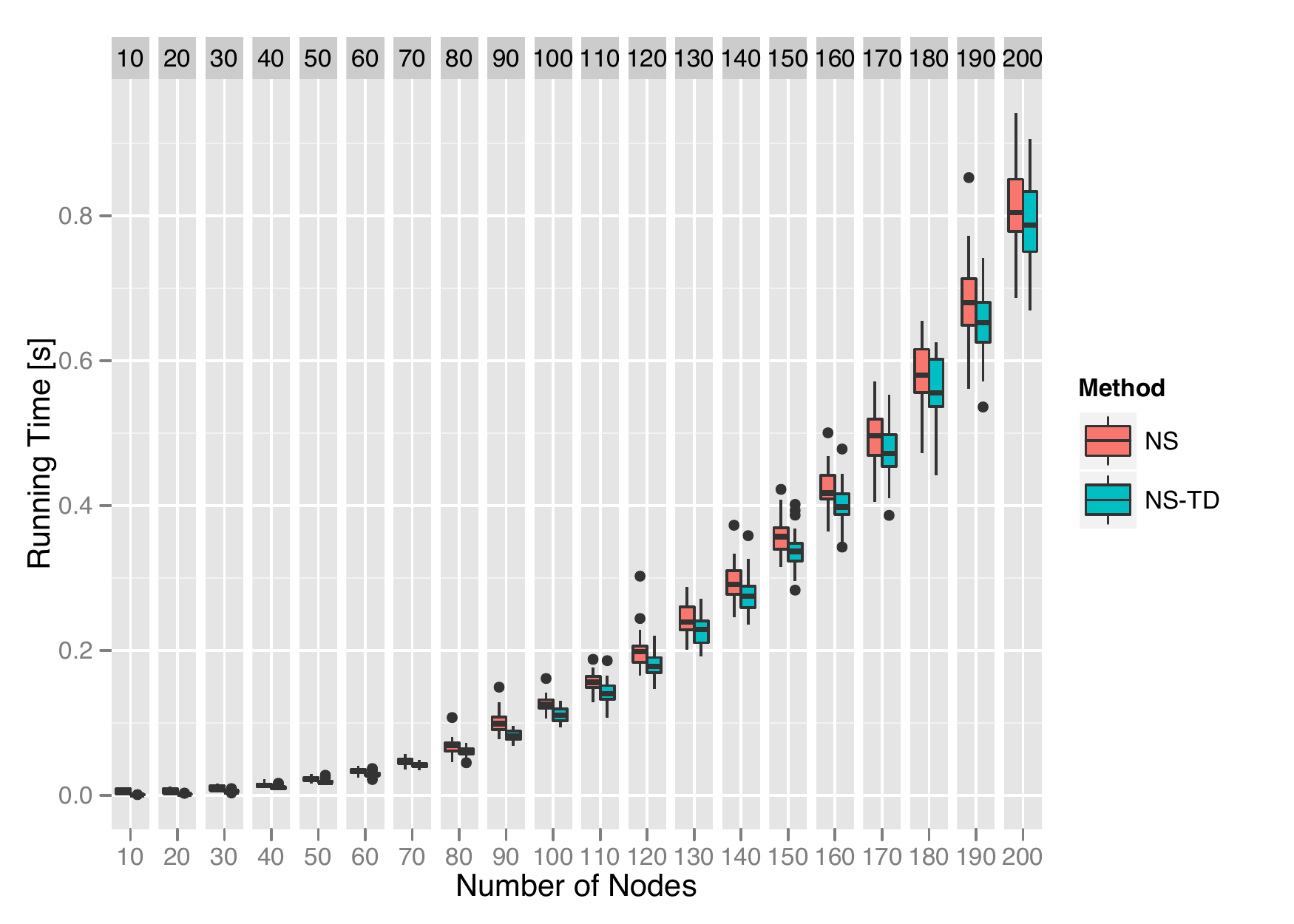}
	}
	\subfloat[\texttt{CompleteK-medium}]{
		\includegraphics[scale=0.51]{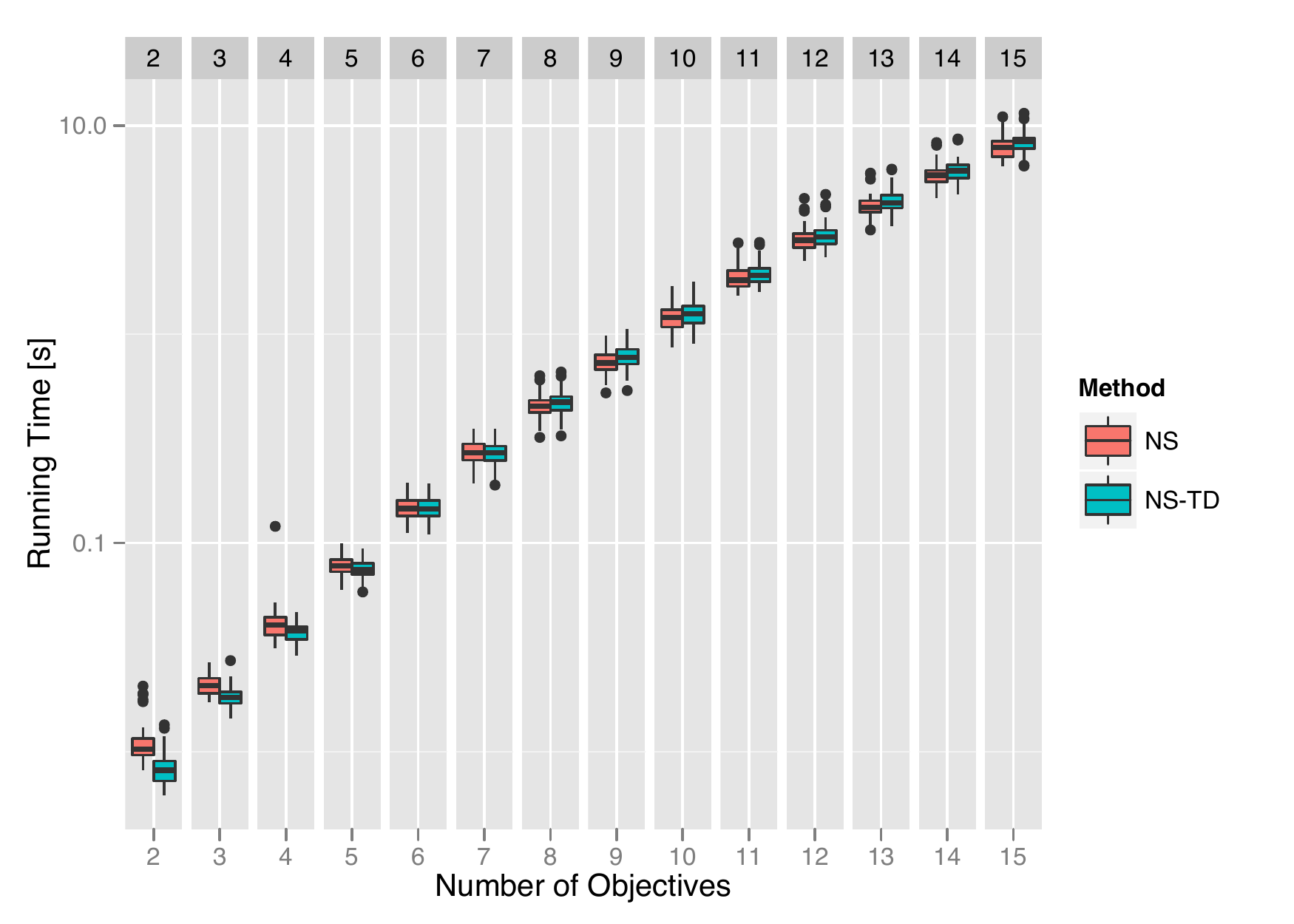}
	}
}\\
\makebox[0cm]{
	\subfloat[\texttt{GridN-medium}]{
		\includegraphics[scale=0.51]{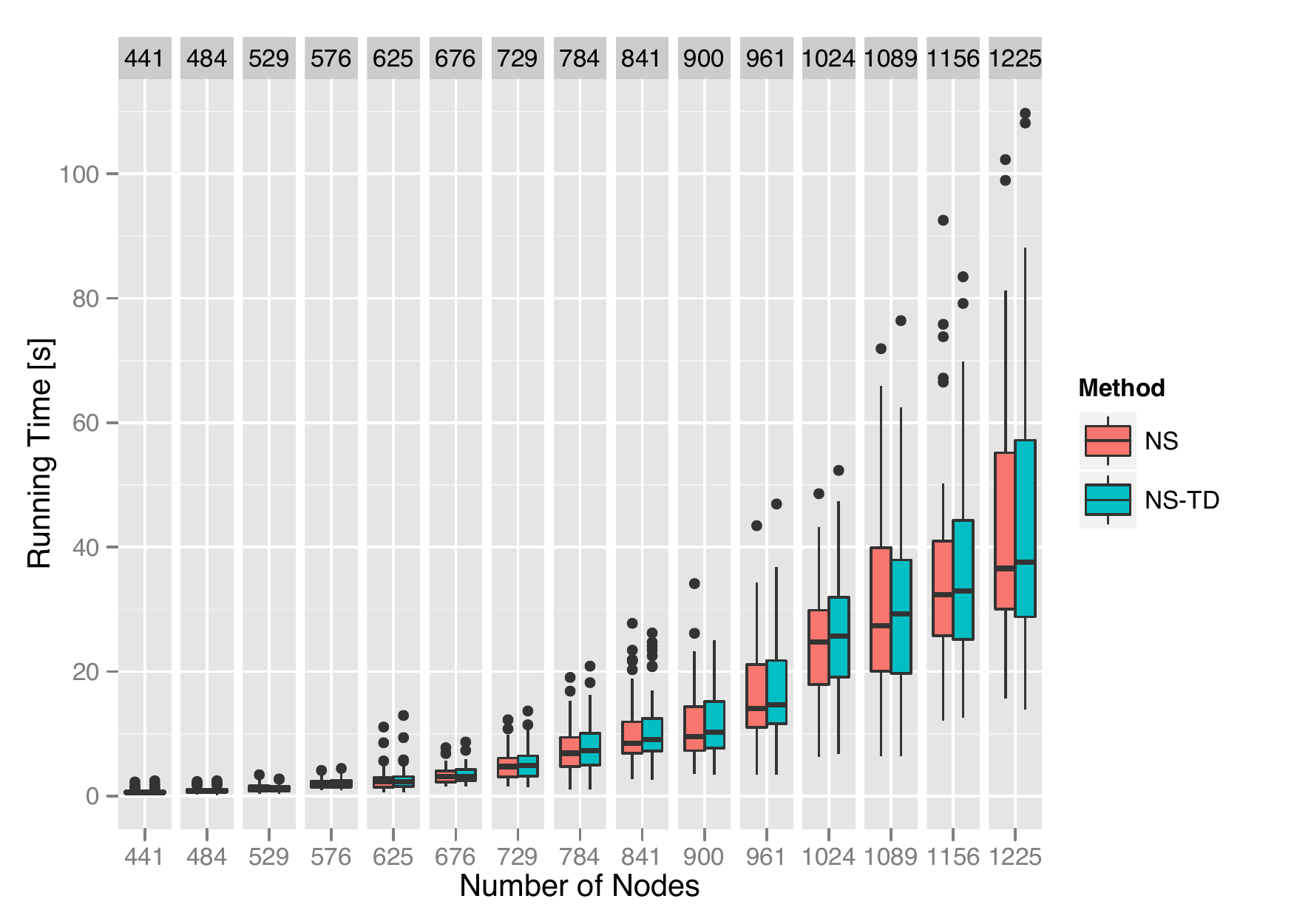}
	}
	\subfloat[\texttt{GridK-medium}]{
		\includegraphics[scale=0.6]{ns-td_GridK-medium}
	}
}\\
\makebox[0cm]{
	\subfloat[\texttt{RandomN-medium}]{
		\includegraphics[scale=0.51]{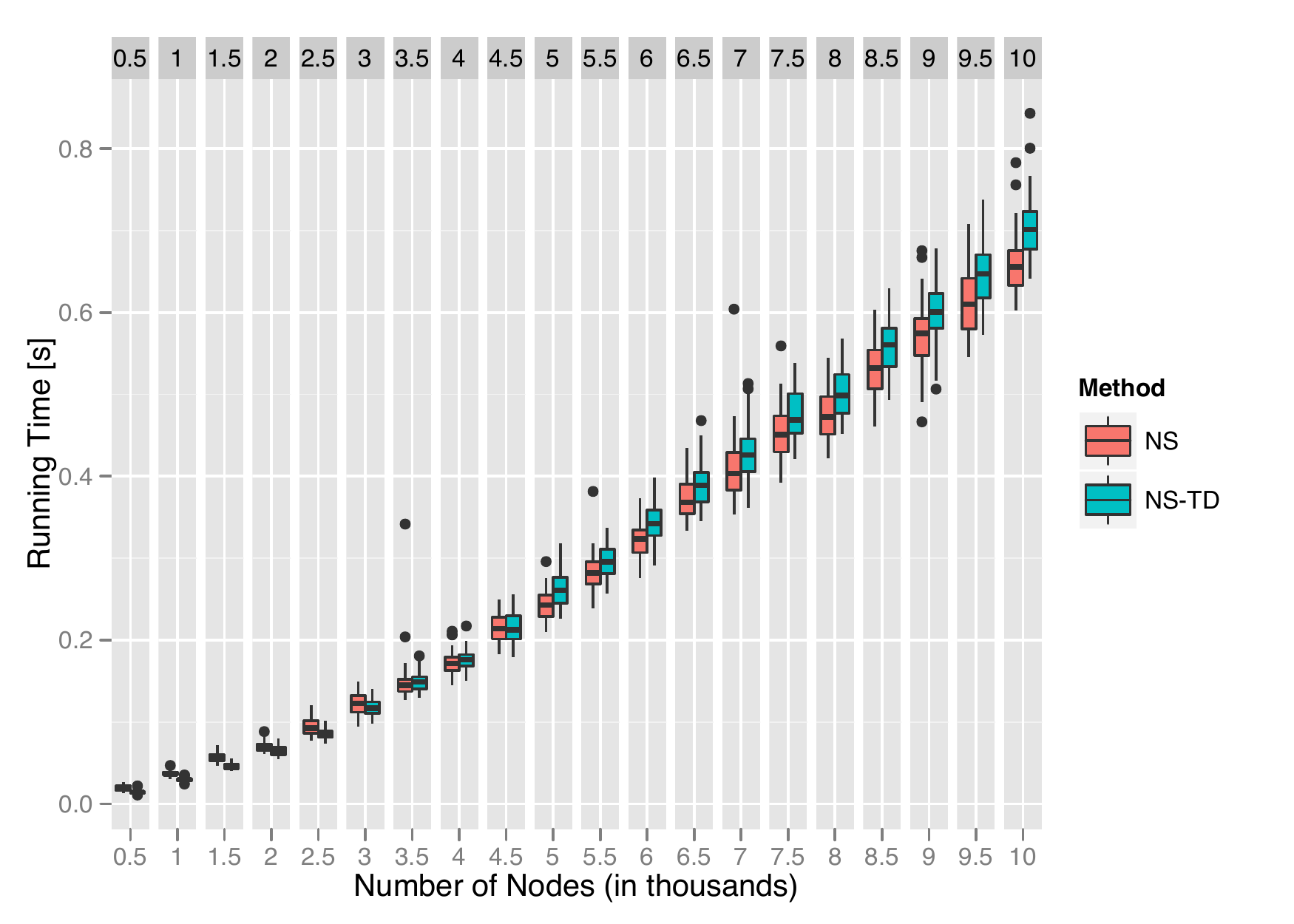}
	}
	\subfloat[\texttt{RandomK-medium}]{
		\includegraphics[scale=0.51]{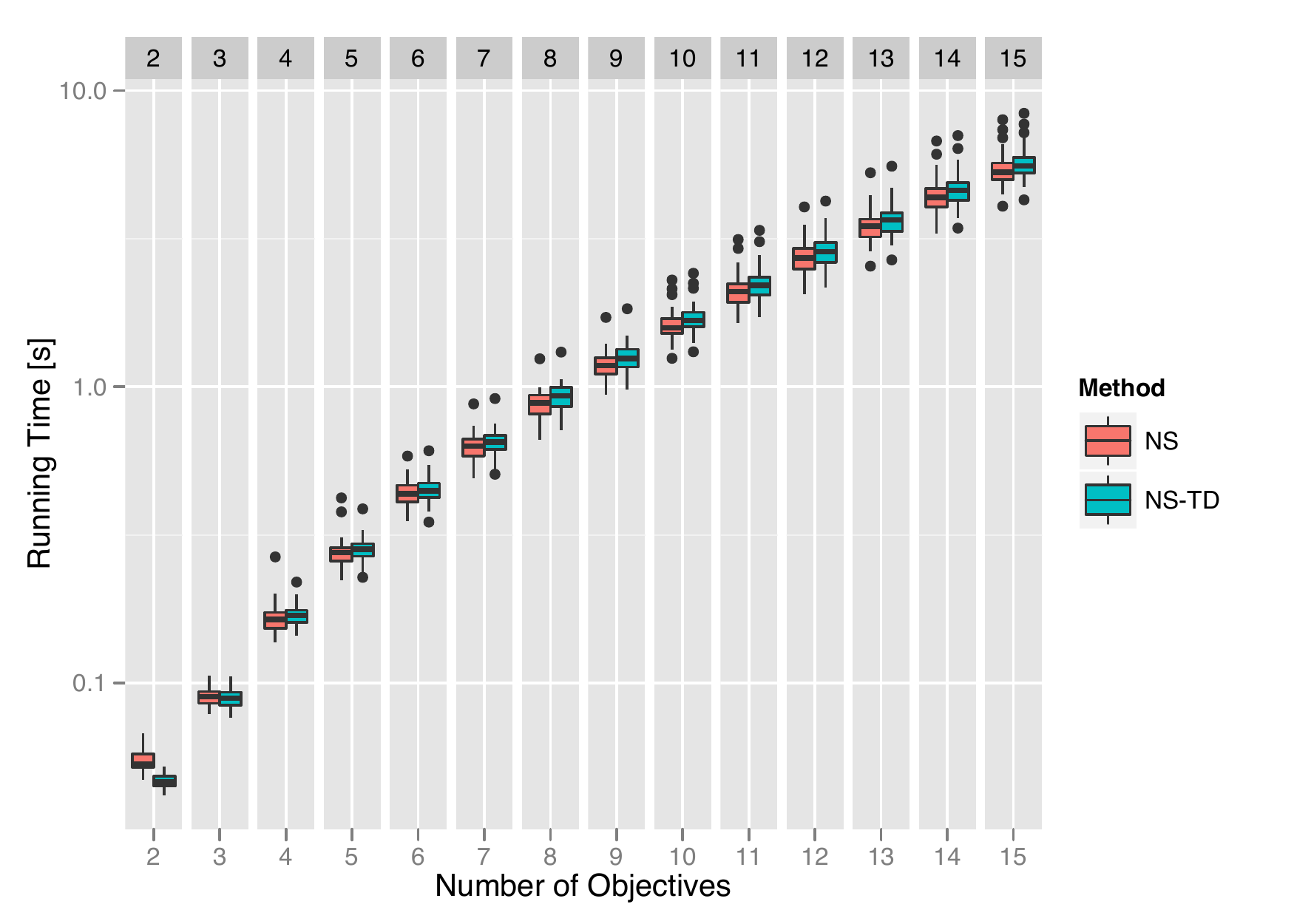}
	}
}
\caption{Comparison of the running times of the node-selection strategy with (NS-TD) and without (NS) tree-deletion pruning}
\label{fig:ns-td:medium}
\end{figure}

\begin{figure}[tb]
\centering
\makebox[0cm]{
	\subfloat[\texttt{CompleteN-large}]{
		\includegraphics[scale=0.6]{ns-td_CompleteN-large}
	}
	\subfloat[\texttt{CompleteK-large}]{
		\includegraphics[scale=0.51]{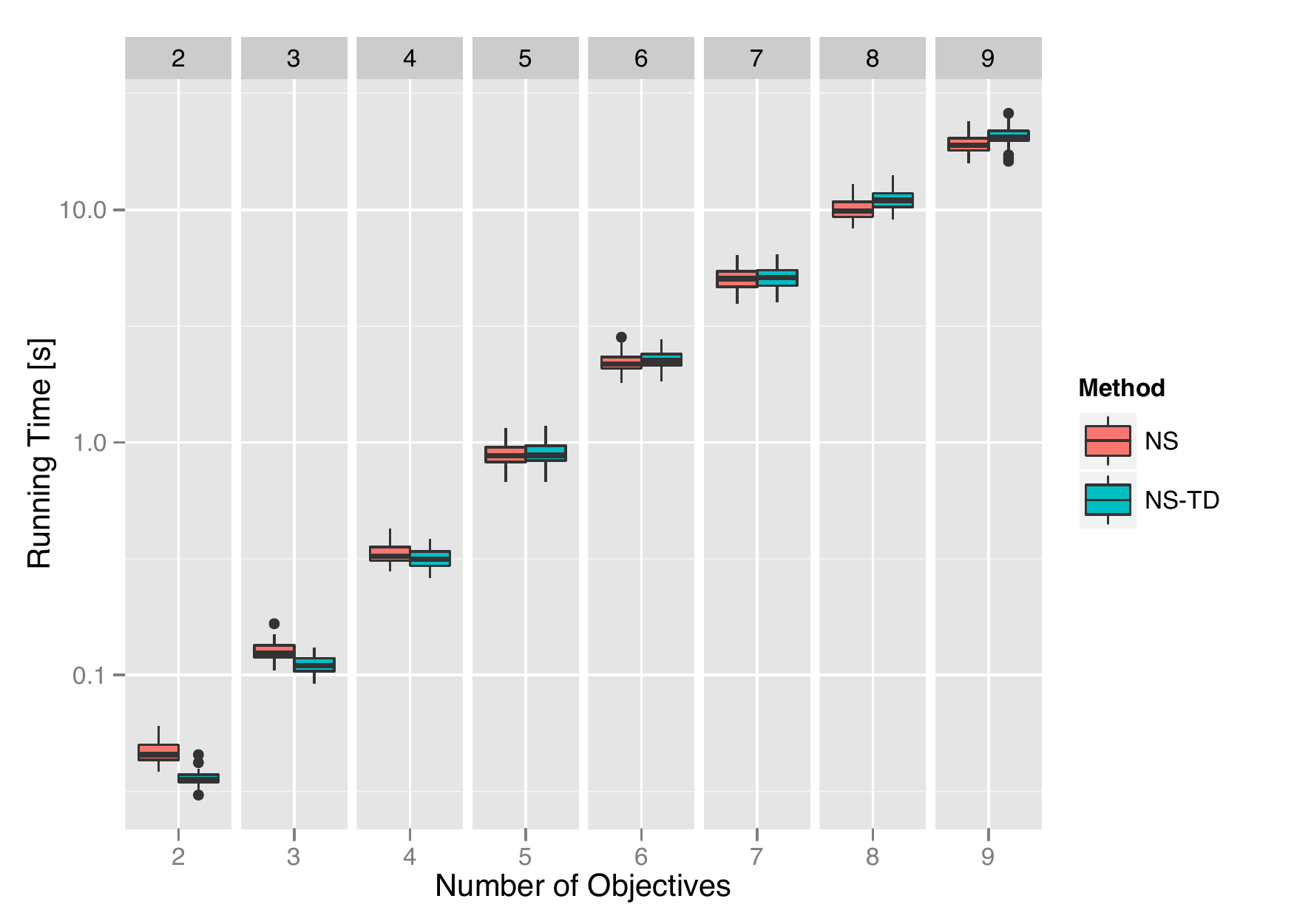}
	}
}\\
\makebox[0cm]{
	\subfloat[\texttt{GridN-large}]{
		\includegraphics[scale=0.51]{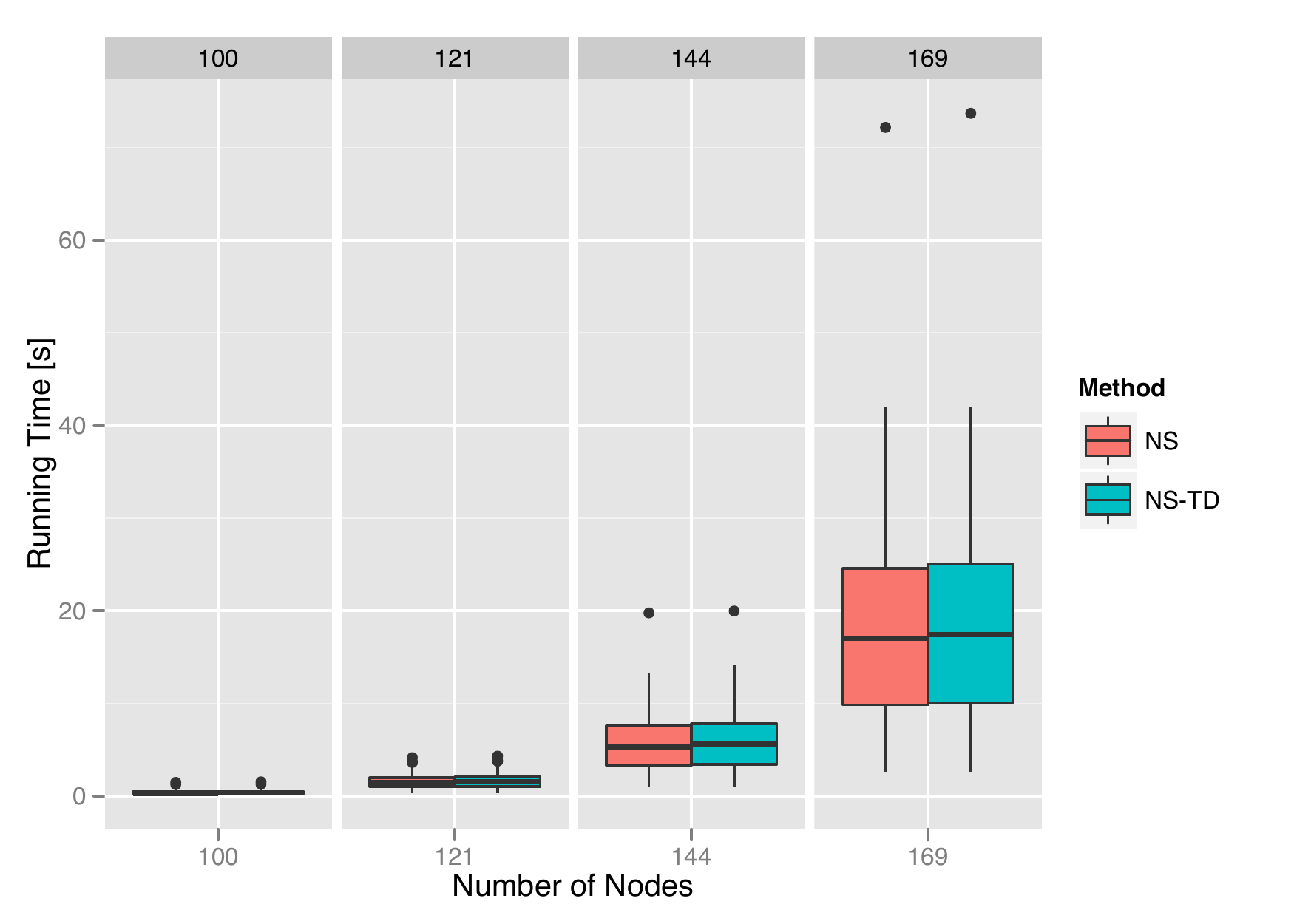}
	}
	\subfloat[\texttt{GridK-large}]{
		\includegraphics[scale=0.51]{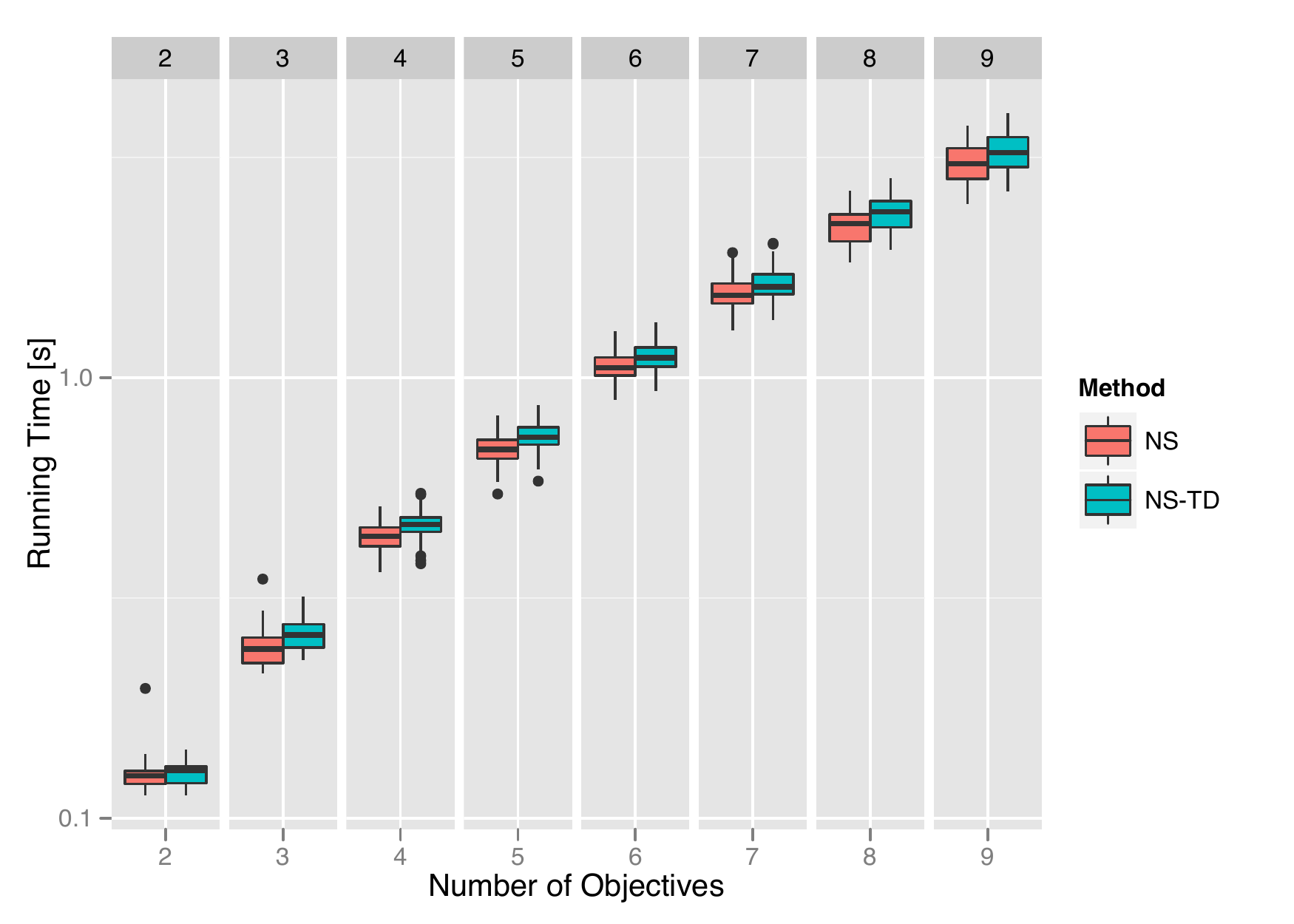}
	}
}\\
\makebox[0cm]{
	\subfloat[\texttt{RandomN-large}]{
		\includegraphics[scale=0.6]{ns-td_RandomN-large}
	}
	\subfloat[\texttt{RandomK-large}]{
		\includegraphics[scale=0.51]{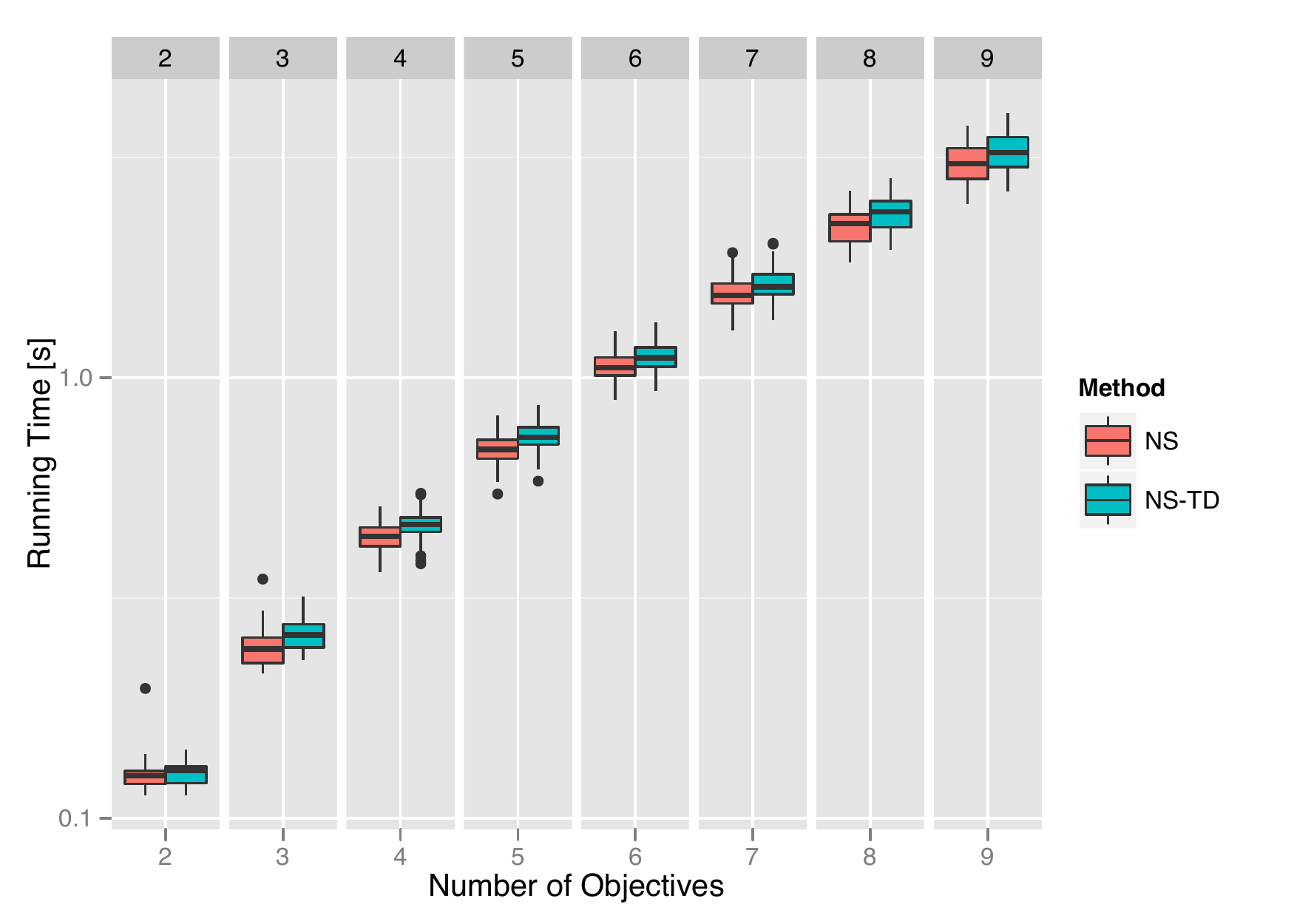}
	}
}
\caption{Comparison of the running times of the node-selection strategy with (NS-TD) and without (NS) tree-deletion pruning}
\label{fig:ns-td:large}
\end{figure}

\begin{figure}[tb]
\centering
\makebox[0cm]{
	\subfloat[\texttt{CompleteN-medium}]{
		\includegraphics[scale=0.51]{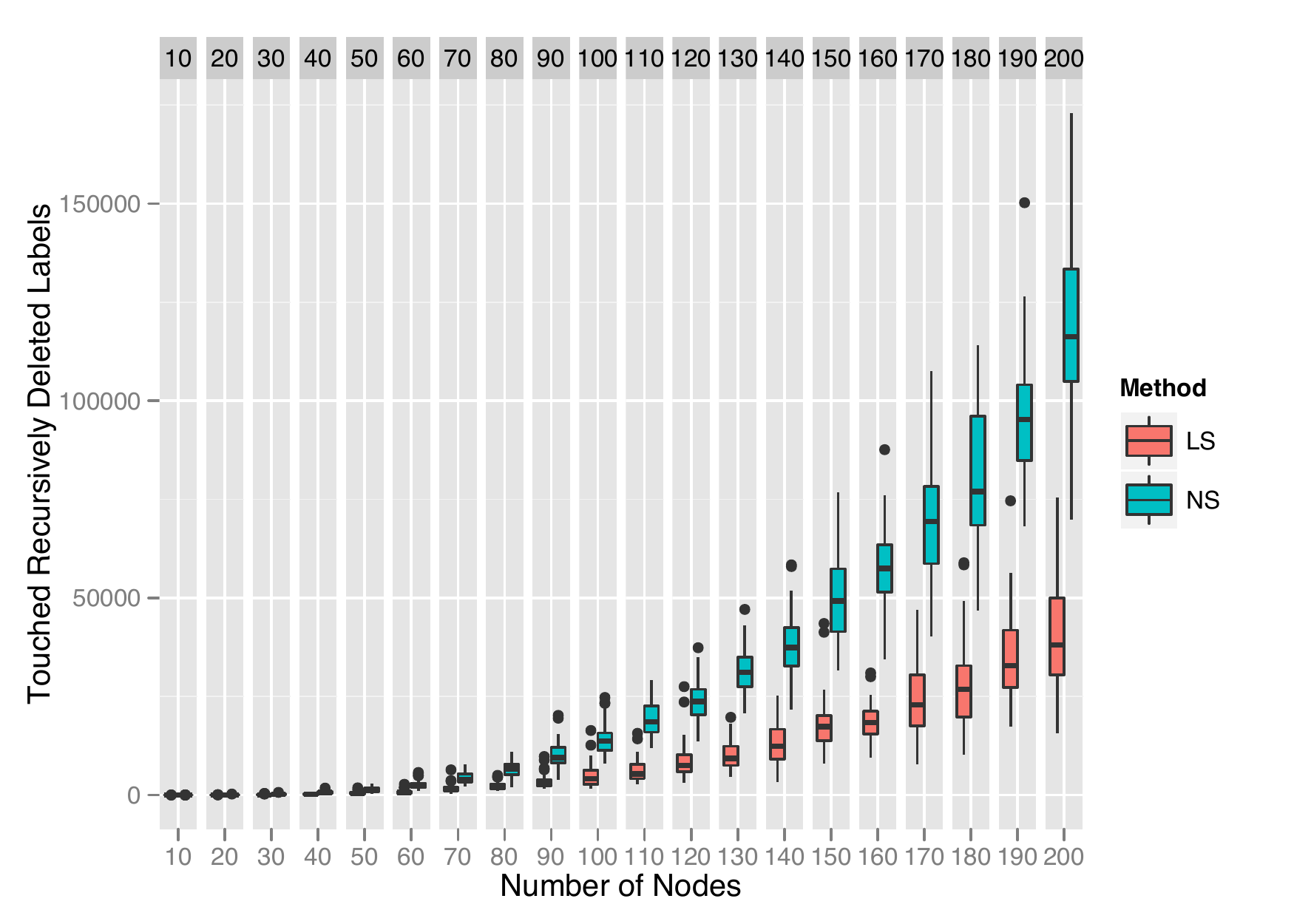}
	}
	\subfloat[\texttt{CompleteK-medium}]{
		\includegraphics[scale=0.51]{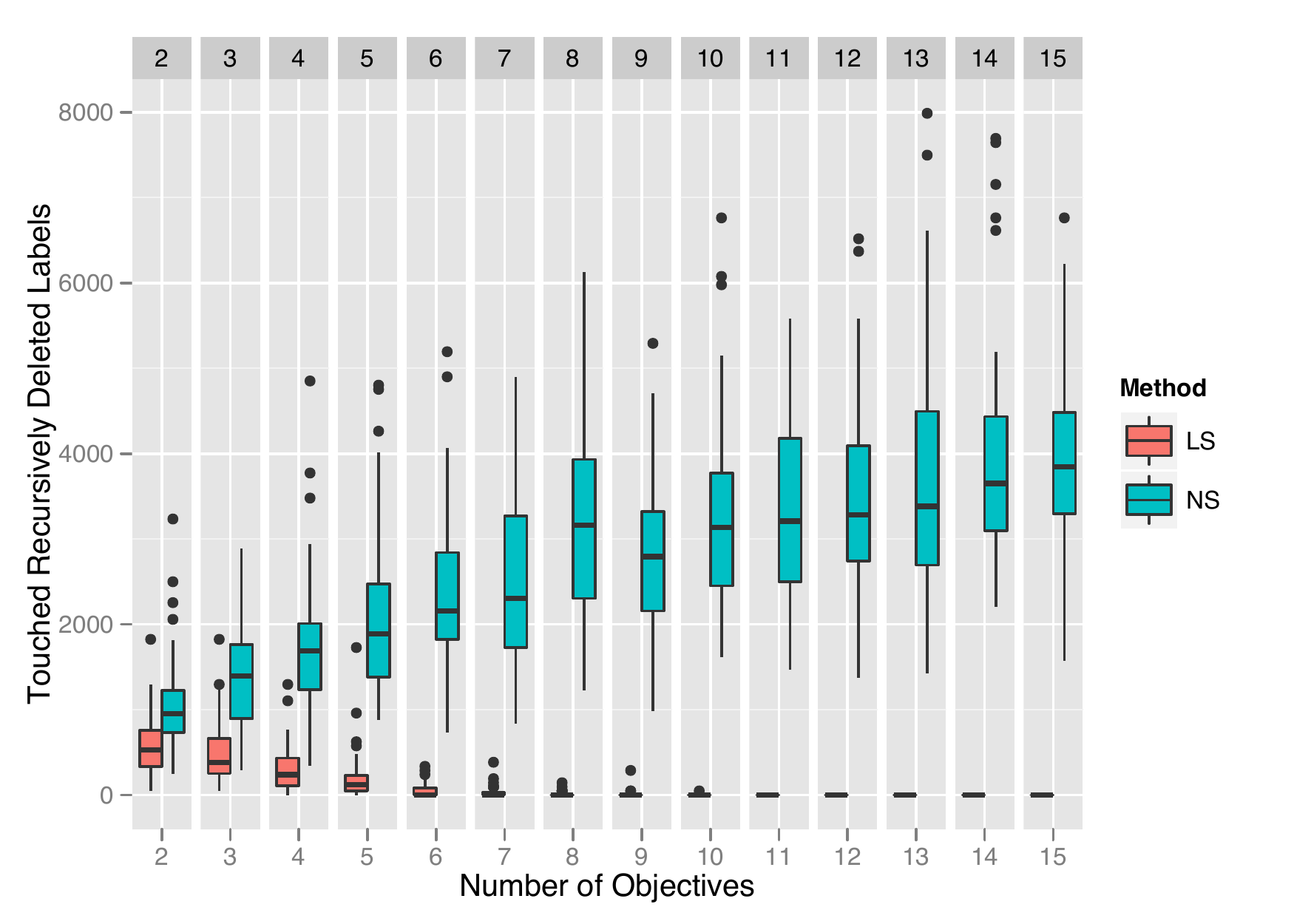}
	}
}\\
\makebox[0cm]{
	\subfloat[\texttt{GridN-medium}]{
		\includegraphics[scale=0.51]{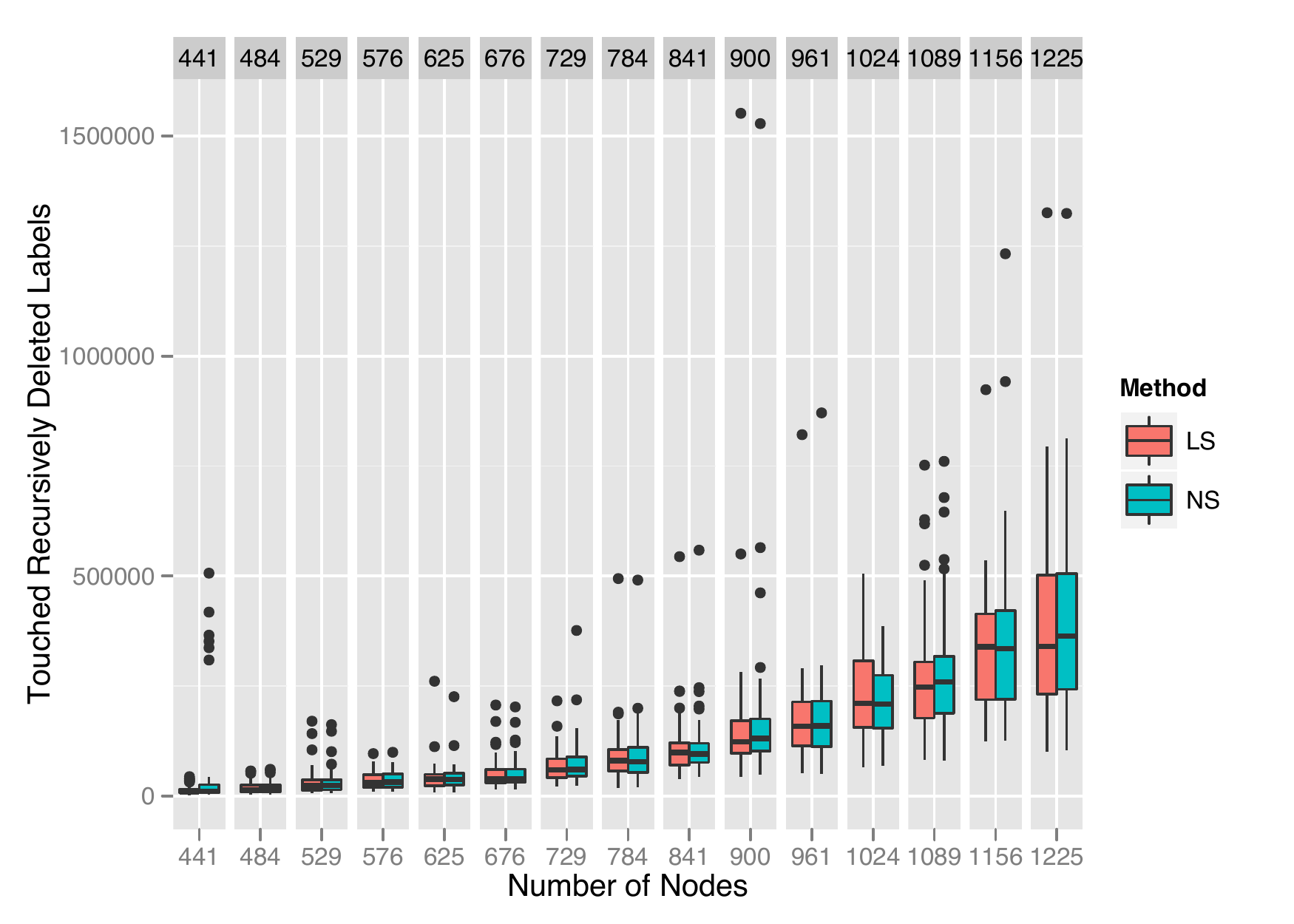}
	}
	\subfloat[\texttt{GridK-medium}]{
		\includegraphics[scale=0.51]{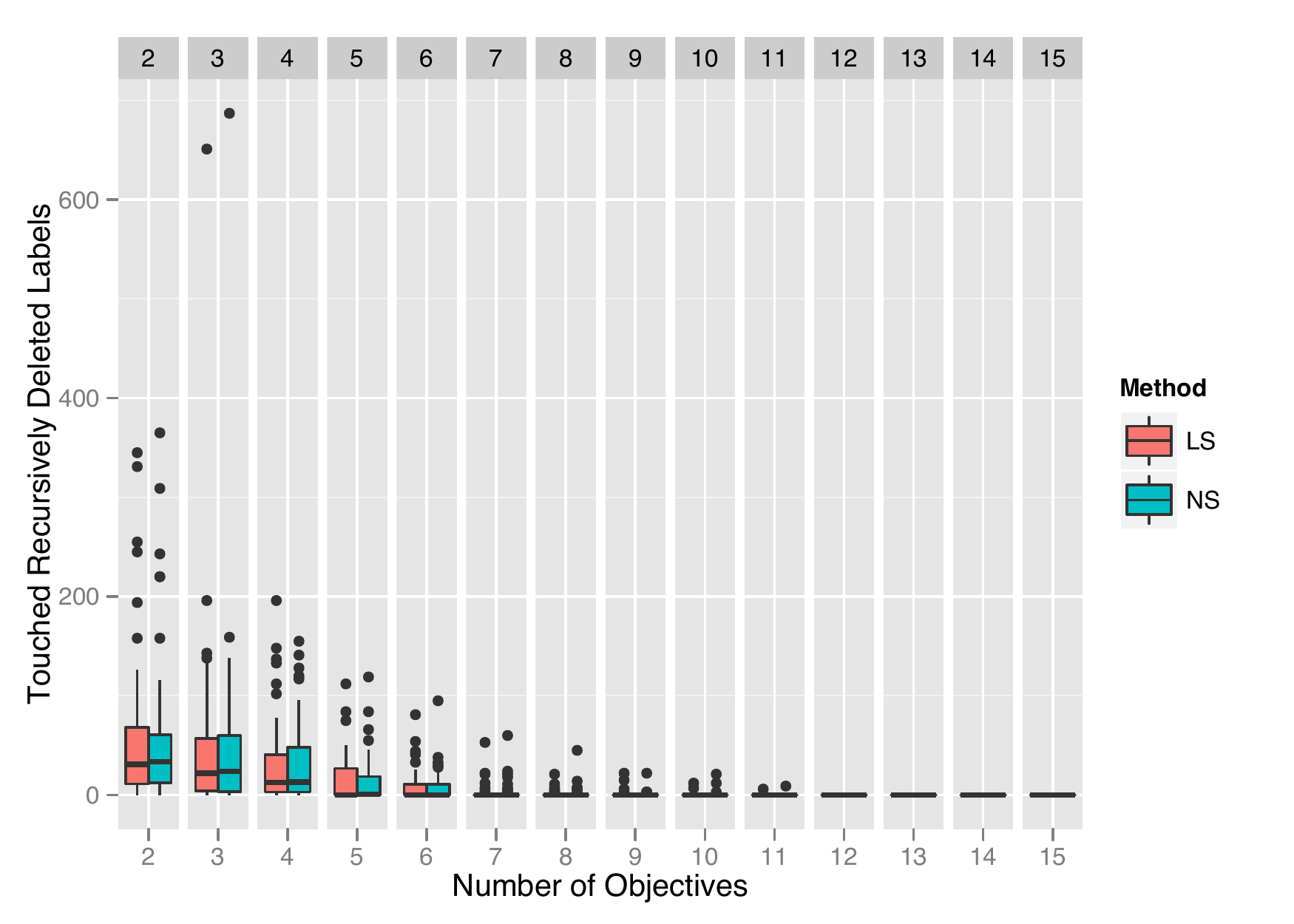}
	}
}\\
\makebox[0cm]{
	\subfloat[\texttt{RandomN-medium}]{
		\includegraphics[scale=0.51]{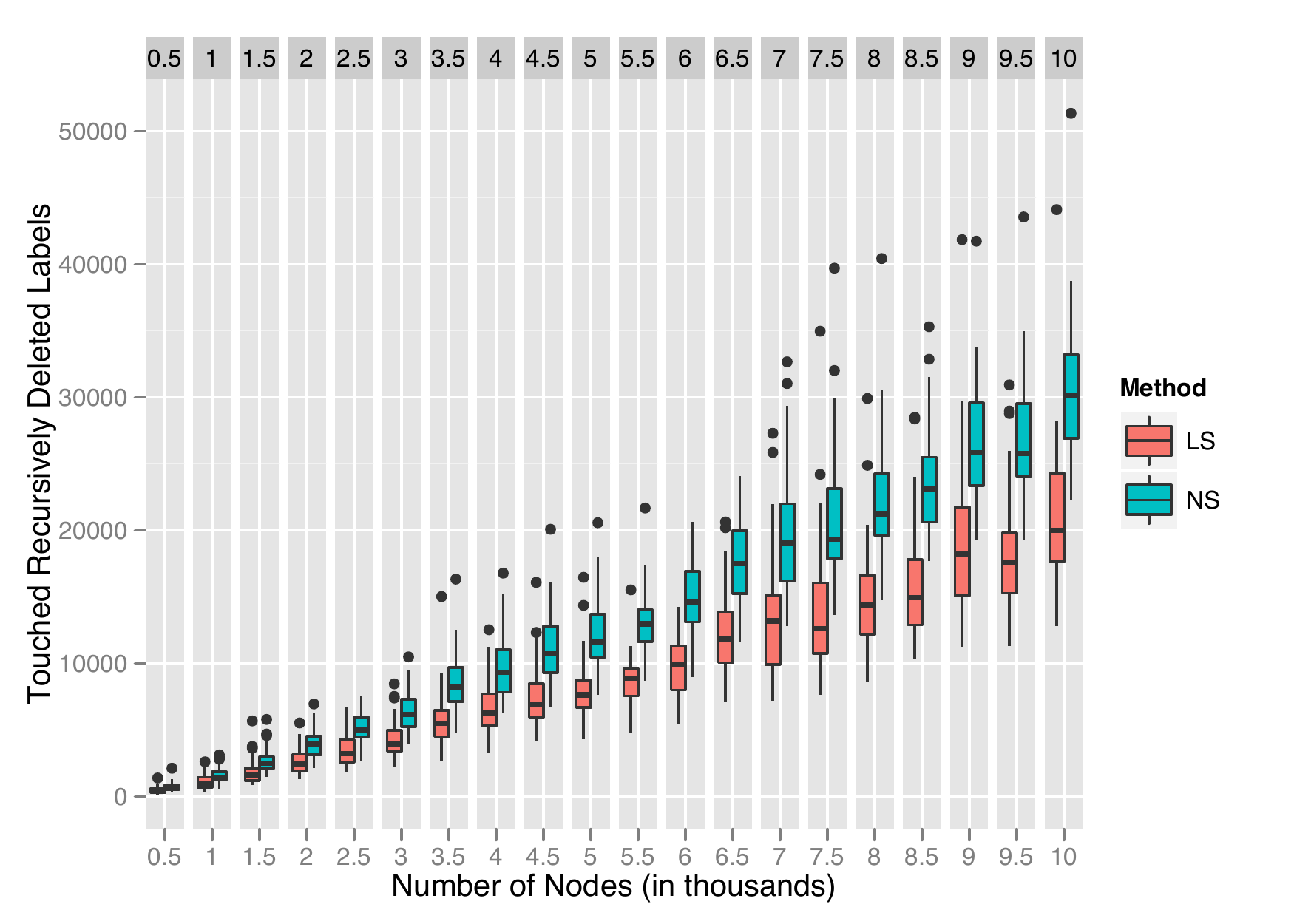}
	}
	\subfloat[\texttt{RandomK-medium}]{
		\includegraphics[scale=0.51]{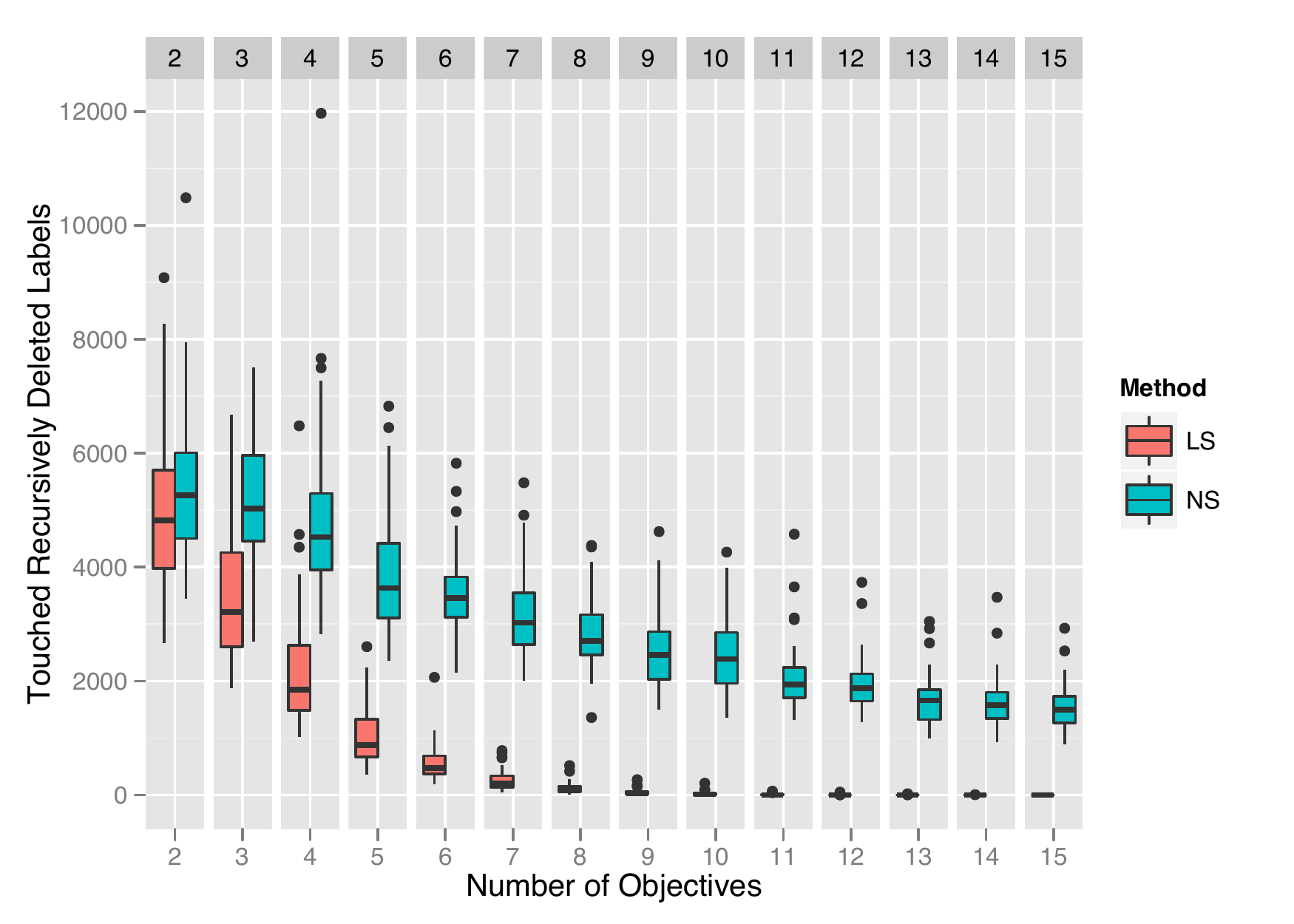}
	}
}
\caption{Measuring how many nodes have been touched which could have been deleted by tree-deletion pruning in the label-selection (LS) and node-selection (NS) strategies}
\label{fig:rec-touch:medium}
\end{figure}

\begin{figure}[tb]
\centering
\makebox[0cm]{
	\subfloat[\texttt{CompleteN-large}]{
		\includegraphics[scale=0.6]{recursive-touch_CompleteN-large}
	}
	\subfloat[\texttt{CompleteK-large}]{
		\includegraphics[scale=0.6]{recursive-touch_CompleteK-large}
	}
}\\
\makebox[0cm]{
	\subfloat[\texttt{GridN-large}]{
		\includegraphics[scale=0.51]{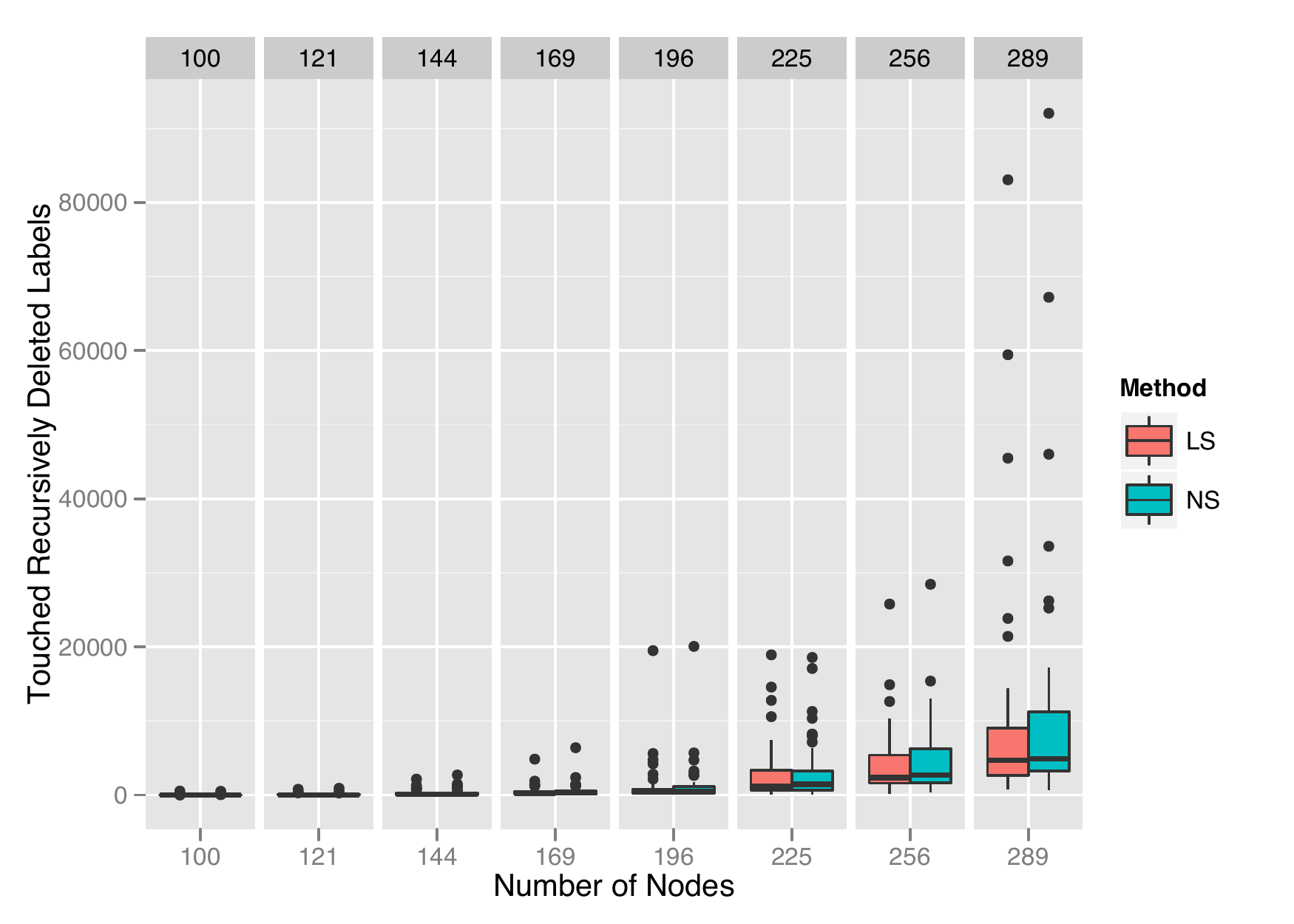}
	}
	\subfloat[\texttt{GridK-large}]{
		\includegraphics[scale=0.6]{recursive-touch_GridK-large}
	}
}\\
\makebox[0cm]{
	\subfloat[\texttt{RandomN-large}]{
		\includegraphics[scale=0.51]{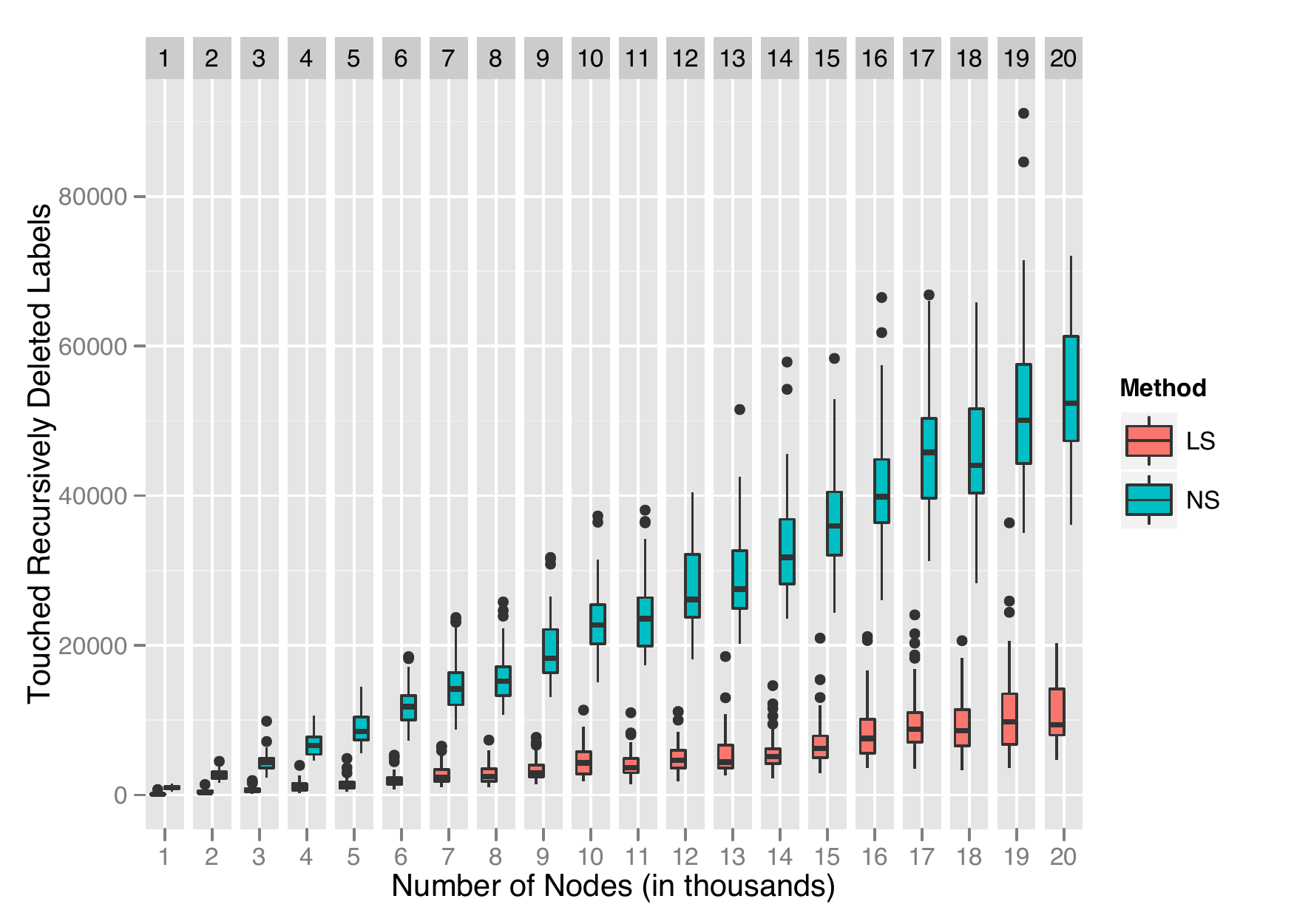}
	}
	\subfloat[\texttt{RandomK-large}]{
		\includegraphics[scale=0.6]{recursive-touch_RandomK-large}
	}
}
\caption{Measuring how many nodes have been touched which could have been deleted by tree-deletion pruning in the label-selection (LS) and node-selection (NS) strategies}
\label{fig:rec-touch:large}
\end{figure}

We provide all boxplots for all experiments in this appendix.
Figs. \ref{fig:ls-ns:medium} and \ref{fig:ls-ns:large} show all the running times on all large and medium sized instances of the artificial instance sets.
Figs. \ref{fig:ns-td:medium} and \ref{fig:ns-td:large} show the comparison of the node-selection strategy with and without TD.
And in Figs. \ref{fig:rec-touch:medium} and \ref{fig:rec-touch:large} we can see the number of labels which could have been delete when we had used TD.
The figures are oversized for better readability and will be removed in a final version of the paper.

\end{document}